\documentclass[pra,reprint,superscriptaddress,bibnotes]{revtex4-1}

\usepackage[T1]{fontenc}
\usepackage[utf8]{inputenc}
\usepackage[australian]{babel}
\usepackage[a4paper,centering,hmargin=1.75cm,vmargin=2cm]{geometry}
\usepackage{amsmath,amssymb,bm}
\usepackage{graphicx,xcolor}
\usepackage{titlesec}
\usepackage{microtype}
\usepackage{braket,siunitx}
\usepackage[version=3]{mhchem}
\usepackage{bbold}
\usepackage{lipsum}
\usepackage{mathtools}
\usepackage{hyperref} 
\usepackage{tabularx}

\graphicspath{{figures/}}
\titleformat*{\paragraph}{\itshape}

\begin{document}

\title{Benchmarking bosonic modes for quantum information with randomized displacements}

\affiliation{School of Physics, University of Sydney, NSW 2006, Australia}
\affiliation{ARC Centre of Excellence for Engineered Quantum Systems, University of Sydney, NSW 2006, Australia}
\affiliation{University of Sydney Nano Institute, University of Sydney, NSW 2006, Australia}

\author{Christophe~H.~Valahu}
\email{christophe.valahu@sydney.edu.au}
\affiliation{School of Physics, University of Sydney, NSW 2006, Australia}
\affiliation{ARC Centre of Excellence for Engineered Quantum Systems, University of Sydney, NSW 2006, Australia}
\affiliation{University of Sydney Nano Institute, University of Sydney, NSW 2006, Australia}

\author{Tomas Navickas}
\thanks{Current address: Q-CTRL, Sydney, NSW 2000, Australia}
\affiliation{School of Physics, University of Sydney, NSW 2006, Australia}
\affiliation{ARC Centre of Excellence for Engineered Quantum Systems, University of Sydney, NSW 2006, Australia}

\author{Michael~J.~Biercuk}
\affiliation{School of Physics, University of Sydney, NSW 2006, Australia}
\affiliation{ARC Centre of Excellence for Engineered Quantum Systems, University of Sydney, NSW 2006, Australia}

\author{Ting Rei Tan}
\affiliation{School of Physics, University of Sydney, NSW 2006, Australia}
\affiliation{ARC Centre of Excellence for Engineered Quantum Systems, University of Sydney, NSW 2006, Australia}
\affiliation{University of Sydney Nano Institute, University of Sydney, NSW 2006, Australia}

\begin{abstract}
Bosonic modes are prevalent in all aspects of quantum information processing. However, existing tools for characterizing the quality, stability, and noise properties of bosonic modes are limited, especially in a driven setting. Here, we propose, demonstrate, and analyze a bosonic randomized benchmarking (BRB) protocol that uses randomized displacements of the bosonic modes in phase space to determine their quality. 
 We investigate the impact of common analytic error models, such as heating and dephasing, on the distribution of outcomes over randomized displacement trajectories in phase space.  We show that analyzing the distinctive behavior of the mean and variance of this distribution - describable as a gamma distribution - enables identification of error processes, and quantitative extraction of error rates and correlations using a minimal number of measurements. We experimentally validate the analytical models by injecting engineered noise into the motional mode of a trapped ion system and performing the bosonic randomized benchmarking protocol, showing good agreement between experiment and theory.  Finally, we investigate the intrinsic error properties in our system, identifying the presence of highly correlated dephasing noise as the dominant process.  
\end{abstract}

\maketitle

\section{Introduction}

Bosonic modes play an important role in various aspects of quantum technologies such as digital quantum computing~\cite{Srensen1999, Knill2001}, continuous-variable quantum computing~\cite{Braunstein2005, Weedbrook2012} and quantum sensing~\cite{Giovannetti2004, Gilmore2021}. Their ubiquity spans diverse physical platforms, such as the normal modes of vibrations in trapped ions~\cite{Bruzewicz2019}, the cavity modes in superconducting circuit Quantum Electrodynamics (cQED)~\cite{Blais2021}, or optical modes in quantum photonics~\cite{Bourassa2021}. In the context of digital quantum computing, error correction codes encoded in bosonic modes can significantly reduce the physical overhead required to implement quantum error correction. Trapped ions and cQED have experimentally demonstrated Gottesman-Kitaev-Preskill (GKP) codes~\cite{Fluhmann2019, CampagneIbarcq2020, deNeeve2022, Matsos2023}, cat codes~\cite{Heeres2017, Gertler2021, lachancequirion2023autonomous} and binomial codes~\cite{Hu2019, Eickbusch2022, Ni2023}. Bosonic modes are also widely used in trapped ions to mediate multi-qubit entangling gates~\cite{Cirac1995, Srensen2000, britton2009}, and quantum control schemes leveraging bosons have demonstrated increased robustness~\cite{Leung2018, Milne2020, Bentley2020}. Bosonic modes can also be used for analog quantum simulations, where encoding information in both the qubit and bosonic modes of a system could solve difficult problems in chemistry and material science~\cite{Gorman2018, MacDonell2021, MacDonell2023, Kang2023, Wang2023, Whitlow2023, Valahu2023, pham2024insitutunable}. Additionally, bosonic modes also allow for unprecedented sensitivities in metrology applications~\cite{biercuknaturenano2011, Wolf2019, McCormick2019, Gilmore2021}.

Given the widespread use of bosonic modes in quantum information, there is a need for quantum characterization, verification and validation (QCVV) tools~\cite{Gheorghiu2018, supic2020, Eisert2020} targeted specifically at bosonic modes, rather than their linked qudit systems. 
The tasks of verification and validation aim to efficiently ensure that an operation acting on a physical system behaves in an intended way. 
In a complementary fashion, characterization aims to gain information about the system, such as Hamiltonian parameters, error rates, and noise mechanisms in order to inform novel strategies for control countermeasures~\cite{Ball_2021} or hardware improvements. 
A particularly prominent QCVV strategy in digital quantum computing is Randomized Benchmarking (RB), commonly used to benchmark average quantum computer hardware performance by measuring the error rate of a gate set via repeated applications of randomized gate sequences \cite{Emerson2005, Magesan2011, Magesan2012, Epstein2014}. RB can also be used to characterize the underlying properties of the limiting error processes~\cite{Ball2016, Keung2016, Mavadia2018, Edmunds2020}. 

Despite the abundance of QCVV protocols for qubit-based systems, only a few solutions are applicable to bosonic modes. 
Several verification and validation protocols have been proposed for analog operations~\cite{Elben2020, Shaffer2023} implementing arbitrary Hamiltonians~\cite{Derbyshire2020, Shaffer2021}, or for bosonic channels~\cite{Wu2019}, but there remains a lack of characterization tools.
Common bosonic characterization protocols involve isolated measurements of decoherence channels, such as measuring the phonon heating rate and the dephasing rate~\cite{Roos2008, Brownnutt2015}. Similar protocols are often used in displacement or phase sensing~\cite{McCormick2019, Wolf2019}.
However, these measurements require the bosonic mode to be idle for a varying amount of time and may not capture the true dynamics during driven interactions that appear in the aforementioned applications.  A step in this direction recently saw successful single-error-channel bosonic-mode noise spectroscopy performed during a quantum logic gate~\cite{Milne2021}.

\begin{figure*}
    \centering
    \includegraphics[width=170mm]{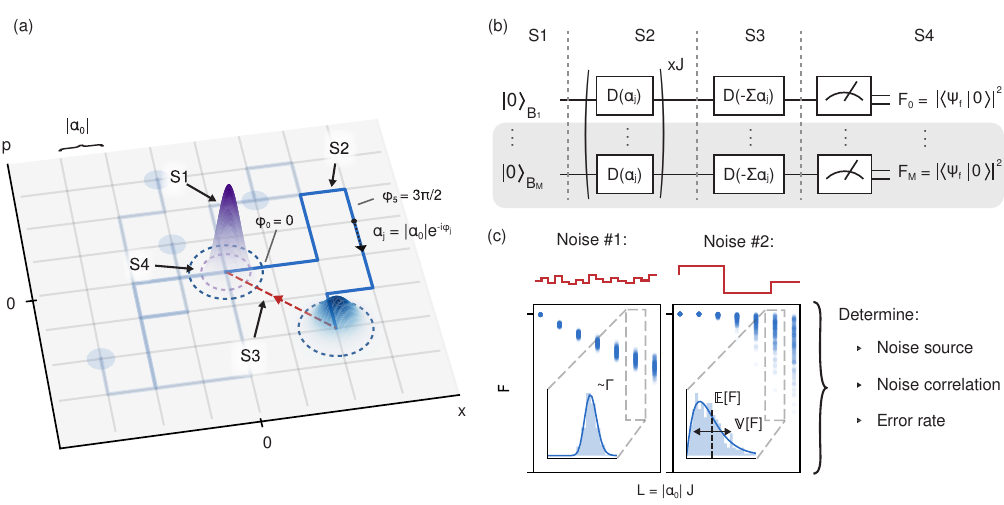}
    \caption{Protocol for randomized benchmarking of bosonic modes. (a) Illustration in phase space of a single bosonic mode undergoing randomized displacements and (b) circuit representation. (S1) The bosonic mode is first initialised to its ground state $\ket{0}_{B_1}$ such that the wavepacket begins at the origin in phase space. (S2) $J$ displacements, each with magnitude of $|\alpha_0|$, are then applied with randomized phases. (S3) A final displacement returns the wavepacket to the origin, (S4) where the fidelity is measured as the overlap between the final wavepacket and $\ket{0}_{B_1}$. Noise during the displacements leads to a larger uncertainty in the wavepacket, and hence a lower fidelity. 
    (c) Experimental measurements of the fidelity for various displacement lengths, $L = |\alpha_0|J$, under two different noise sources with distinct characteristics: (left) low correlations and small amplitude from heating, and (right) high correlations and large amplitude from dephasing. Probability distributions of the fidelities are well-approximated by $\Gamma$-distributions (see inset), which are uniquely described by the mean, $\mathbb{E}[\mathcal{F}]$, and variance, $\mathbb{V}[\mathcal{F}]$. The different noise sources result in two different functional forms of the decay of $\mathbb{E}[\mathcal{F}]$ over $L$ (left and right blue markers). The higher amplitude of the second noise source leads to a faster decay rate of $\mathbb{E}[\mathcal{F}]$. High correlations lead to a skewness in the distribution (right inset). From few measurements of $\mathbb{E}[\mathcal{F}]$ and $\mathbb{V}[\mathcal{F}]$ at varying $L$, one can determine the noise source, its correlation and an error rate.}
    \label{fig:protocol}
\end{figure*}

In this work, we address the lack of generalized QCVV tools tailored to driven bosonic modes by introducing a novel characterization and benchmarking method: bosonic randomized benchmarking (BRB). Analogous to conventional RB, this process involves the application of a sequence of randomized displacements in phase space, followed by an inversion step and a measurement.  Following previous efforts~\cite {Ball2016, Mavadia2018, Edmunds2020}, we define quantitative measures that allow a user to infer detailed information about noise processes and error correlations from the distribution of outcomes over different randomized sequences. In particular, we show that measuring the mean and variance of a gamma distribution enables the identification and discrimination of various common noise sources in hardware.  We analytically derive the corresponding properties for motional heating, rapidly fluctuating dephasing, and strongly correlated dephasing, revealing distinct behaviors that are then validated in experiments using engineered noise in a trapped-ion device.  In addition, we use the BRB protocol to characterize the intrinsic limiting noise processes in our hardware and show that data are consistent with a slowly fluctuating quasistatic dephasing process.

\section{Protocol}

The bosonic randomized benchmarking (BRB) protocol tests and characterizes the quality of bosonic modes in a quantum system by repeatedly applying random displacements of the mode in two-dimensional phase space, as shown in Fig.~\ref{fig:protocol}.  These random displacements accumulate according to a predefined pattern corresponding to a single sequence, and the bosonic mode's net displacement is measured against the expected displacement under perfect manipulation.  This process is then repeated after reinitialization using a new random sequence of displacements, and the average fidelity of the applied sequences is calculated (illustrated in Fig.~\ref{fig:protocol}(c)). The decay of this average fidelity with the length of the random displacement sequence is extracted by varying the number of displacements in a sequence. This routine is expressed in the following formal steps:

\begin{itemize}
    \item[S1.] Initialize the bosonic mode B$_\mathrm{i}$ to its ground state, $\ket{\psi_{\mathrm{B}_i}} = \ket{0}$.
    \item[S2.] Apply $J$ sequential displacements, where the $j^{th}$ displacement is described by $\hat{\mathcal{D}}(\alpha_j) = e^{-i(\alpha_j \hat{a} + \alpha_j^* \hat{a}^\dagger)}$, with $\alpha_j = |\alpha_0|e^{-i \phi_j}$. The phases $\phi_j$ are randomly sampled with uniform weights from $\phi_j \in \{0, \frac{\pi}{2}, \pi, \frac{3\pi}{2} \}$, and $|\alpha_0|$ is the unit length of the displacements.  The phases of this step are constrained to a discrete set for simplicity; they can instead be randomly sampled from the range $\phi_j \in [0, 2\pi]$ without impact.
    \item[S3.] Apply a final displacement $\hat{\mathcal{D}}(-\sum_j\alpha_j)$ to return the state
    to the ground state.
    \item[S4.] Measure the fidelity as the overlap of the final state with the ground state, $\mathcal{F}_i = | \langle\psi_{\mathrm{B}_i, \mathrm{f}}| 0\rangle |^2$. 
    \item[S5.] Repeat steps S1-S4 $M$ times to obtain a noise averaged value $\tilde{\mathcal{F}}$. 
    \item[S6.] Repeat steps S1-S5 $N$ times for different circuit realisations with $J$ displacements, $(\alpha_0, \alpha_1, ..., \alpha_J)$, where each $\alpha_j$ is randomly sampled as in S2.
\end{itemize}

This approach can be directly applied to the characterization of either a single or multiple bosonic modes (see Fig.~\ref{fig:protocol}(b)). To characterize multiple modes, the $j^{th}$ displacement is applied simultaneously to all modes. The fidelity of each mode, $\mathcal{F}_i$, is then recorded at the completion of the circuit. 

Characterization of BRB outputs for either a single or multiple bosonic modes is conducted by analyzing the functional behavior of two simple metrics linked to the distribution of fidelity outcomes over randomizations at fixed $L = |\alpha_0| J$: the mean, $\mathbb{E}[\mathcal{F}]$, and variance, $\mathbb{V}[\mathcal{F}]$ (see Fig.~\ref{fig:protocol}c). 
From conventional RB, it is known that the average error of a gate set can be inferred from the decay of $\mathbb{E}[\mathcal{F}]$ as a function of the sequence length $L$, and variability of the decay's form can be linked to the type of error process~\cite{Fogarty2015, FigueroaRomero2021}.  Moving beyond the decay behavior of the mean, the mean and variance of the distribution of fidelities entirely define a distribution at fixed sequence length well described by a $\Gamma$-distribution, whose properties are intimately linked to the detailed characteristics of the underlying error process (this follows similar insights first introduced for RB applied to qubits in Refs.~\cite{Ball2016, Mavadia2018}).  
As a general observation, rapidly fluctuating (uncorrelated) noise processes are manifested as a gamma distribution well-approximated by a Gaussian centered around $\mathbb{E}[\mathcal{F}]$, while noise processes exhibiting strong temporal correlations produce a highly asymmetric and broad distribution.  

Therefore, the experimentally simple protocol and measurement of a few quantitative parameters allows one to benchmark the performance of a driven bosonic mode, and gain insight into the dominant noise process and its temporal correlation properties.  In the following section, we analytically derive the expected form of the $\Gamma$-distribution for key noise processes encountered in typical experimental settings.

\section{Analytical model of error manifestation in BRB}
\label{sec:error_model}

We derive an error model for three distinct noise sources that often dominate bosonic interactions: \textit{heating}, \textit{Markovian (uncorrelated) dephasing}, and \textit{DC (correlated) dephasing}. 
Heating of the bosonic modes is modelled by random photon/phonon gains and losses, which are implemented by the creation operator, $\hat{a}^\dagger$, and annihilation operator, $\hat{a}$, respectively. 
The Hamiltonian $\epsilon(t)\hat{a}^\dagger \hat{a}$ models dephasing, where the correlation of $\epsilon(t)$ determines whether the dynamics are Markovian or non-Markovian. We parametrize the Markovianity by the correlation length, $\mathcal{M}_n$, defined as the number of displacements $n$ over which the noise is assumed to be constant~\cite{Ball2016}.

We begin by expressing the impact of each of the aforementioned noise sources on the form of the distribution over randomizations output by BRB in general terms.  From here onwards, we drop the subscript $i$ that labels the $i$th bosonic mode B$_i$; furthermore, without loss of generality, we consider a single mode.  At the end of a sequence of randomized displacements, the final state is (ignoring a global phase)
\begin{equation} \label{eq:final_state}
    \ket{\psi_\mathrm{B,f}} = \hat{\mathcal{D}}(-\sum_{j=0}^{J-1} \alpha_j)\prod_{j=0}^{J-1} \hat{\mathcal{D}}(\tilde{\alpha}_j) \ket{0} = \hat{\mathcal{D}}(\alpha_\epsilon)\ket{0}.
\end{equation}
Here, $\prod_{j=0}^{J-1} \hat{\mathcal{D}}(\tilde{\alpha}_j)$ correspond to the $J$ randomized displacements, while $\hat{\mathcal{D}}(-\sum_{j=0}^{J-1} \alpha_j)$ is the final single-step displacement which ideally returns the bosonic mode to the origin. Displacements of $\tilde\alpha$ ($\alpha$) designate noisy (ideal) displacements. We note that, in the following derivations, we assume the error in the final reversal pulse to be negligible; the validity of this assumption increases in the limit of large $J$. 

Using the expression of Eq.~\ref{eq:final_state}, we can derive an analytical expression for the fidelity as
\begin{equation} \label{eq:fidelity_base}
   \mathcal{F} = e^{-|\alpha_\epsilon|^2}.
\end{equation}
where $\alpha_\epsilon$ represents the overall parasitic displacement from the phase space origin due to noise. This parasitic displacement is derived from Eq.~\ref{eq:final_state} to give
\begin{equation} \label{eq:alpha_epsilon}
    \alpha_\epsilon = \tilde{\alpha}_\mathrm{tot} - \alpha_\mathrm{tot},
\end{equation}
where $\alpha_\mathrm{tot}$ and $\tilde{\alpha}_\mathrm{tot}$ are the ideal and noisy total displacements, respectively. 
The displacement $\hat{\mathcal{D}}(\alpha_\mathrm{tot})$ is implemented by applying a noise-free control Hamiltonian,
\begin{equation} \label{eq:control_hamiltonian}
    H_\mathrm{c} = \frac{\Omega}{2}\hat{a}e^{-i \phi_\mathrm{c}(t)} + \mathrm{h.c.},
\end{equation}
where $\Omega$ is the interaction strength of the displacement drive and $\phi_\mathrm{c}(t)$ is a piecewise constant function that encodes the randomized displacements of the BRB protocol. The step size of a single displacement of the BRB protocol is set by $|\alpha_0| = \Omega \Delta \tau /2$, where $\Delta\tau$ is the single-step duration. From $H_\mathrm{c}$, the ideal total trajectory becomes
\begin{align} \label{eq:noisy_alpha_tot}
    \alpha_\mathrm{tot} = & - \frac{i\Omega J \Delta\tau}{2} \sum_{j=0}^{J-1}  e^{- i \phi_j}.
\end{align}
The analytical form of the noisy trajectory, $\tilde{\alpha}_\mathrm{tot}$, will depend on the exact error mechanism under consideration. In the following section, we derive several expressions for $\tilde{\alpha}_\mathrm{tot}$ and $\alpha_\epsilon$ under heating and dephasing errors. We also consider amplitude and phase fluctuations in Appendix~\ref{app:dephasing_noise_model}.

With the generalized expression of Eq.~\ref{eq:alpha_epsilon} of the resulting parasitic displacement, $\alpha_\epsilon$, it becomes possible to analytically calculate the noise averaged fidelity, $\tilde{\mathcal{F}} = \langle \mathcal{F} \rangle_M$, taken in the limit of many noise realisations, $M \rightarrow \infty$.  We can then calculate the distribution of outcomes over randomizations and the statistical moments of this distribution:  $\mathbb{E}[\tilde{\mathcal{F}}]$, and variances, $\mathbb{V}[\tilde{\mathcal{F}}]$.  

In all cases, we find that the distributions of the noise-averaged fidelities are well-approximated by $\Gamma$-distributions, $\Gamma(a, b)$, where the shape, $a$, and rate, $b$, are uniquely defined by the statistical moments above with  $a = \mathbb{E}^2/\mathbb{V}$ and $b = \mathbb{V} / \mathbb{E}$. We derive an expression for the mean and variance under each noise process in the following subsection. We also calculate a decay constant, $\eta$, which describes the rate at which the mean fidelity decreases with $L$.

The key parameters that describe the fidelity distributions and behaviours under heating, Markovian dephasing and DC dephasing are summarized in Table~\ref{tab:error_model}. Each noise mechanism has a unique mean and variance, and as a result exhibits a different $\Gamma$-distribution over randomizations. Therefore, from a single set of experimentally measured fidelities over different randomizations and sequence lengths, one can deduce the dominant noise mechanism, its error rate, and the noise's correlation.

\subsection{Heating}

We first consider errors induced by heating, where the bosonic mode is coupled to a thermal bath. We model heating as random kicks in phase space that displace the bosonic mode~\cite{James1998b, rasmusson2024}. This is captured by setting $\tilde{\alpha}_j = \alpha_j + \epsilon_j$ in Eq.~\ref{eq:final_state}, where $\epsilon_j \sim \mathcal{N}_\mathcal{C}(0, \sigma_\mathrm{h}^2)$ and $\mathcal{N}_\mathcal{C}$ denotes a complex normal distribution. $\sigma_\mathrm{h}^2$ is the variance of the random kicks due to heating. We further assume that all $\epsilon_j$ are independent and identically distributed and have correlation length $\mathcal{M}_n = 1$. The parasitic displacement of Eq.~\ref{eq:alpha_epsilon} becomes $\alpha_\epsilon = \sum_{j=0}^{J-1} \epsilon_j $, where, using the central limit theorem, $\alpha_\epsilon \sim \mathcal{N}_\mathcal{C}(0, J \sigma_\mathrm{h}^2)$. The mean of the noise-averaged fidelity is then calculated from Eq.~\ref{eq:fidelity_base} (see Appendix~\ref{app:heating_noise_model} for details), resulting in
\begin{align} \label{eq:mean_var_heating}
    & \mathbb{E}[\tilde{\mathcal{F}}] = \frac{1}{1 + \eta_\mathrm{h} L},
\end{align}
where the error rate is 
\begin{equation}
    \eta_\mathrm{h}  = \frac{2 \sigma^2_\mathrm{h}}{\Omega \Delta\tau}.
\end{equation}
It may be more convenient, however, to express the variance as $\sigma^2_\mathrm{h} = \gamma_\mathrm{h} \Delta\tau$, where $\gamma_\mathrm{h}$ is a heating rate, in units of $\mathrm{quanta} / s$, and represents the number of phonons gained per second~\cite{Brownnutt2015}. This is an experimentally measurable quantity in trapped ions and other experimental systems that is often used as a figure of merit. With this, the error rate becomes
\begin{equation} \label{eq:error_rate_heating}
    \eta_\mathrm{h} = 2 \gamma_\mathrm{h}/\Omega.
\end{equation}

The variance, $\mathbb{V}$, of the fidelity under heating is
\begin{align} \label{eq:var_heating}
    \mathbb{V}[\tilde{\mathcal{F}}] \propto \mathcal{O}(M^{-1}).
\end{align}
Since the heating noise process commutes with the displacements of the BRB protcol, the noise-averaged fidelity is independent of the randomized trajectory. Therefore, $\mathbb{V}$ is dominated by quantum projection noise and scales inversely with the number of noise measurements, $M$. In the limit of many measurements, $M \rightarrow \infty$, the variance reduces to zero. 

The resulting shape and rate parameters of the $\Gamma$-distribution are found from the mean of Eq.~\ref{eq:mean_var_heating} and the variance of Eq.~\ref{eq:var_heating}, resulting in $a \propto \mathcal{O}(M)$ and $b \propto \mathcal{O}(M^{-1})$.  Calculated expressions are summarized in Table~\ref{tab:error_model}.

\subsection{Dephasing}
We now turn our attention to the effects of dephasing, which are modelled by the noisy Hamiltonian $H_\mathrm{dephasing} = \epsilon_\delta(t) \hat{a}^\dagger \hat{a}$. With this, the control Hamiltonian of Eq.~\ref{eq:control_hamiltonian} becomes
\begin{equation} \label{eq:control_hamiltonian_dephasing}
    \tilde{H}_\mathrm{c} = \frac{\Omega}{2}\hat{a}e^{-i (\phi_\mathrm{c}(t) + \epsilon_\delta(t) t)} + \mathrm{h.c.},
\end{equation}
which contains additional frequency fluctuations captured by $\epsilon_\delta(t)$. The resulting noisy total trajectory is 
\begin{align} \label{eq:noisy_alpha_tot}
    \tilde{\alpha}_\mathrm{tot} = & - \frac{i\Omega}{2} \sum_{j=0}^{J-1}  e^{- i \phi_j} \nonumber \int_{j\Delta \tau}^{(j+1)\Delta \tau} e^{- i \epsilon_\delta(t) t} dt,
\end{align}
and the parasitic displacement of Eq.~\ref{eq:alpha_epsilon} evaluates to
\begin{equation} \label{eq:alpha_epsilon_dephasing}
    \alpha_{\epsilon} = - \frac{i\Omega}{2} \sum_{j=0}^{J-1} e^{-i \phi_j} \int^{(j+1)\Delta\tau}_{j \Delta\tau} dt  (e^{-i \epsilon_\delta(t) t} - 1).
\end{equation}

For Markovian (uncorrelated) noise, the frequency fluctuations have a correlation length $\mathcal{M}_n = 1$ and we replace $\epsilon_\delta(t) \rightarrow \epsilon_{\delta, j}$ in Eq.~\ref{eq:alpha_epsilon_dephasing}. The corresponding autocorrelation function is $\langle\epsilon_{\delta, i}\epsilon_{\delta, j}\rangle = \sigma_\delta^2 \delta(i - j)$, where $\sigma_\delta^2$ is the variance, $\delta(\cdot)$ is the Dirac delta function and $\epsilon_{\delta, j} \sim \mathcal{N}(0, \sigma_\delta^2)$ is normally distributed. For quasi-static DC (correlated) dephasing noise, the frequency fluctuations have a correlation length $\mathcal{M}_n = J$, and we set $\epsilon_\delta(t) \rightarrow \epsilon_\delta$ in Eq.~\ref{eq:alpha_epsilon_dephasing}. $\epsilon_\delta \sim \mathcal{N}(0, \sigma_\delta^2)$ is normally distributed and has variance $\langle \epsilon_\delta^2 \rangle = \sigma_\delta^2$.

The mean fidelity for both Markovian and DC dephasing noise is evaluated to be (see detailed derivations in Appendix~\ref{app:dephasing_noise_model}), 
\begin{align} \label{eq:fidelity_dephasing}
    & \mathbb{E}[\tilde{\mathcal{F}}] = \frac{1}{1 + (\eta_\mathrm{d} L)^3},
\end{align}
where the dephasing error rate is 
\begin{align} \label{eq:error_rate_dephasing}
    \eta_\mathrm{d} = \left(\frac{ 4 |\alpha_0|  \sigma_\delta^2}{3 \Omega^2}\right)^{1/3}.
\end{align}

\def\arraystretch{2}%
\setlength{\tabcolsep}{5pt}
\begin{table}[]
\begin{tabular}{ccc}
\hline \hline
         & Heating & Dephasing \\ \hline 
$\mathbb{E}$ &  $(1 + \eta L)^{-1}$  & $(1 + (\eta L)^3)^{-1}$ \\ 
$\mathbb{V}$ &   $\mathcal{O}(M^{-1})$   & $\mathcal{C} \mathbb{E}(1 - \mathbb{E})^2/ (2 - \mathbb{E})$ \\ 
$\eta$ & $ 2 \gamma_\mathrm{h}/\Omega $ & $(4 |\alpha_0|\sigma_\delta^2/3\Omega^2)^{1/3}$  \\
$a$ & $\mathcal{O}(M)$ & $\mathcal{C}^{-1}\mathbb{E}(2-\mathbb{E})/(1 - \mathbb{E})^2$ \\ 
 $b$ & $\mathcal{O}(M^{-1})$ & $\mathcal{C} (1 - \mathbb{E})^2/(2 - \mathbb{E})$ \\
 \hline \hline
\end{tabular}
\caption{\label{tab:error_model} Summary of means, $\mathbb{E}$, and variances, $\mathbb{V}$, of noise-averaged fidelities, $\tilde{\mathcal{F}}$. $\eta$ is the decay rate of the means. $a$ and $b$ are the shape and rate parameters of a Gamma distribution, $\Gamma(a, b)$, that approximates the probability distribution of the noise averaged fidelities. $L = |\alpha_0| J$ is the distance travelled by the randomised displacements. The variance, rate and shape parameters under dephasing are parametrized by $\mathcal{C}$, which depends on the noise's correlation. We find $\mathcal{C}_\mathrm{Mar.} = 0.071$ and $\mathcal{C}_\mathrm{DC} = 0.572$ for Markovian and DC noise, respectively. All other symbols are defined in the main text.}
\end{table}

\begin{figure*}[t]
    \centering
    \includegraphics{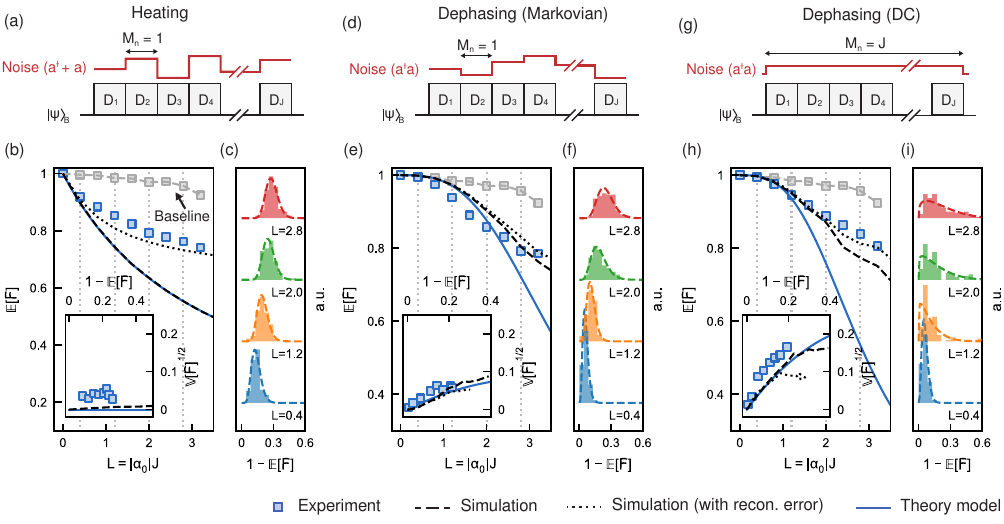}
    \caption{Experimental validation of theoretical error models, where we consider heating (a-c), Markovian (uncorrelated) dephasing (d-f) and DC (correlated) dephasing (g-i). 
    (a, d, g) The correlations of the noise are defined by the correlation length $M_n$. 
    (b, e, i) Experimentally measured fidelities are compared to numerical simulations (dashed and dotted lines) obtained from Eq.~\ref{eq:noisy_alpha_tot} and theory models of Table.~\ref{tab:error_model} (solid line) for varying sequence lengths, $L = |\alpha_0| J$, with $\alpha_0 = 0.1$. Simulations with reconstruction errors consider the effects of the imperfect measurement protocol (see Appendix \ref{app:reconstruction_measurements}). A baseline measurement (grey squares) is taken in the absence of noise injection to ensure that the measured decays are not limited by background noise. 
    Engineered noise is set such that $\gamma_\mathrm{h} = \SI{1.53(16)}{phonons / ms}$ for heating noise, $\sigma_\delta/2\pi = \SI{600}{Hz}$ for Markovian dephasing and $\sigma_\delta/2\pi =  \SI{900}{Hz}$ for DC dephasing, resulting in error rates $\eta_\mathrm{h} = 0.29$, $\eta_\mathrm{d} = 0.26$ (Markovian) and $\eta_\mathrm{d} = 0.34$ (DC), respectively. These were chosen such that the resulting decays are not limited by the system's noise floor. 
    To qualitatively compare theory and experiment, we normalize the experimentally measured means by their value at $L=0$, $\mathbb{E} \approx 0.92$, and offset the variances under noise injection (inset) by the baseline variance. Histograms of the fidelity measurements are plotted for lengths $L = \{0.4, 1.2, 2.0, 2.8\}$, indicated by the grey vertical dotted lines. 
    (c, f, i) The experimental histograms are compared to ideal Gamma probability distributions $\Gamma(a, b)$ (dashed lines), whose shape and rate parameters $a$ and $b$ are directly calculated from the experimentally measured mean and variance and have no free fitting parameters.}
    \label{fig:exp_validation} 
\end{figure*}

The variance of the fidelity under dephasing is 
\begin{align} \label{eq:variance_dephasing}
    \mathbb{V}[\tilde{\mathcal{F}}] & = \mathcal{C} \left( \frac{1}{1 + 2(\eta L)^3} - \frac{1}{(1 + (\eta L)^3)^2} \right) \nonumber \\ 
    & = \mathcal{C} (1-\mathbb{E}[\tilde{\mathcal{F}}])^2 \frac{\mathbb{E}[\tilde{\mathcal{F}}]}{2 - \mathbb{E}[\tilde{\mathcal{F}}]}.
\end{align}
The variance between Markovian and DC noise only differs by a constant, $\mathcal{C}$, in Eq.~\ref{eq:variance_dephasing}. $\mathcal{C}$ can be determined by fitting numerical simulations (see Appendix~\ref{app:dephasing_noise_model} for details); we find $\mathcal{C}_\mathrm{Mar.} = 0.071$ for Markovian noise and $\mathcal{C}_\mathrm{DC} = 0.572$ for DC noise. Therefore, the variance leads to a non-zero value for Markovian and DC noises, even in the limit of infinite noise averaging. 

The behaviour of the mean fidelities under dephasing is similar for both qubit randomized benchmarking (RB) and bosonic randomized benchmarking (BRB), with mean fidelities for Markovian and DC noise being approximately equal~\cite{Ball2016, Edmunds2020}. However, the variance of fidelity under dephasing differs: in qubit RB, the variance is non-zero in the limit of infinite averaging under DC noise and approaches zero under Markovian noise~\cite{Ball2016, Edmunds2020}, while in BRB, the variance remains non-zero for both DC and Markovian noise. The discrepency in behaviour under Markovian noise arises from the different noise susceptibility of randomised sequences between qubits and bosonic modes. In qubit RB, the noise susceptibility for Markovian noise is independent of the randomized benchmarking sequence (see Refs.~\cite{Ball2016, Edmunds2020} for an in-depth discussion). However, the noise susceptibility of bosonic modes is correlated with the bosonic randomized sequence, since the effects of dephasing are intimately linked to the distance of the mode in phase space, even for Markovian noise. Displacements that cause large excursions in phase space populate higher Fock states that are more susceptible to dephasing~\cite{McCormick2019}.

\section{Experimental demonstration}
\subsection{Experimental system}
We experimentally validate and then deploy the BRB protocol for bosonic-noise characterization and quantification using a single $^{171}\ce{Yb}^+$ ion confined in a macroscopic Paul trap. The bosonic mode under test is encoded in one of the ion's radial motional modes with an oscillation frequency of $\omega_r/2\pi = \SI{1.33}{MHz}$. An associated qubit is encoded in the $^2\ce{S}_{1/2}$ hyperfine ground state separated by $\omega_0/2\pi = \SI{12.6}{GHz}$, where we assign the labels $\ket{\downarrow}_\mathrm{Q} \equiv \ket{F=0, m_f=0}$ and $\ket{\uparrow}_\mathrm{Q} \equiv \ket{F=1, m_f=0}$. A detailed description of the system and experimental methods can be found in Ref.~\cite{MacDonell2023, Valahu2023}. 

In this system, the qubit plays an essential role in bosonic readout, allowing transduction from the bosonic modes of the ion's motion to a subsystem with a strong projective measurement. Interactions with the qubit and the bosonic modes are driven by Raman transitions that are enacted by a $\SI{355}{nm}$ pulsed laser. Driving transitions at frequencies $\omega_r = \omega_0 - \omega_r$ and $\omega_b = \omega_0 + \omega_r$ leads to red (r)- and blue (b)-sideband interactions with Hamiltonians in the frame of the qubit and bosonic mode given by \cite{Wineland1998},
\begin{align}
    & H_\mathrm{r} = \frac{\Omega}{2} \hat{\sigma}^+ \hat{a} e^{- i \phi_\mathrm{r}} + \mathrm{h.c.},  \\
    & H_\mathrm{b} = \frac{\Omega}{2} \hat{\sigma}^+ \hat{a}^\dagger e^{- i \phi_\mathrm{b}} + \mathrm{h.c.},
\end{align}
where $\hat{\sigma}^+ = \ket{\uparrow}\bra{\downarrow}$. The Rabi frequency, $\Omega/2\pi \approx \SI{1.68}{kHz}$, and phases, $\phi_\mathrm{r,b}$, are set by the Raman laser's amplitude and phase. 

Displacements of the bosonic mode are implemented by simultaneously driving the red- and blue-sideband transitions, resulting in the state-dependent force (SDF) Hamiltonian
\begin{equation} \label{eq:sdf_hamiltonian}
    H_\textrm{SDF} = \frac{\Omega}{2}\hat{\sigma}_x \hat{a} e^{-i \phi} + \textrm{h.c.}.
\end{equation}
Applying $H_\mathrm{SDF}$ for a duration $t$ enacts displacements conditional on the qubit state, $\hat{\mathcal{D}}(\hat{\sigma}_x \alpha)$, where $\alpha = \Omega t e^{-i \phi}/2$ and $\phi = \phi_\mathrm{r} = - \phi_\mathrm{b}$ is set by varying the relative phases of the red- and blue-sideband fields. In practice, the phases $\phi_\mathrm{r, b}$ are varied by modulating the radio-frequency signal produced by a direct digital synthesizer that drives an acousto-optic modulator in the path of one of the Raman beams. The displacements $\hat{\mathcal{D}}(\alpha)$ in the BRB protocol are implemented with state-dependent forces; this approach is validated through numerics to be indistinguishable from alternate state-independent displacements obtained from Eq.~\ref{eq:control_hamiltonian} (see Appendix~\ref{app:displacements}).

The fidelity of Eq.~\ref{eq:fidelity_base} is experimentally retrieved by measuring the qubit state probability after mapping information from the bosonic mode to the qubit by applying a $\pi$-pulse on the red-sideband. We find that, for small errors $|\alpha_\epsilon|^2 \ll 1$, the measured qubit probability approximates the noise-averaged fidelity, $\tilde{\mathcal{F}}$ (see Appendix~\ref{app:reconstruction_measurements}). 

\begin{figure}
    \centering
    \includegraphics[]{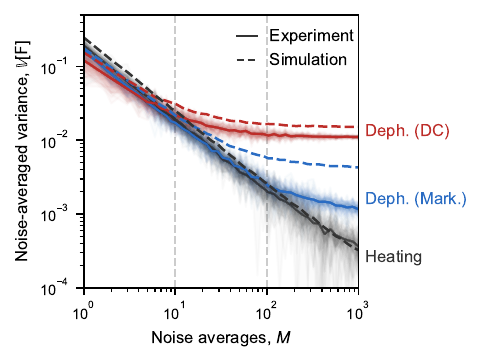}
    \caption{Noise-averaged variance of the fidelity, $\mathbb{V}[\tilde{\mathcal{F}}]$, with increasing noise averages, $M$. Dephasing (DC and Markovian) and heating noise are engineered with the same parameters as Fig.~\ref{fig:exp_validation}. $N = 50$ circuit repetitions are performed with $J=32$ randomized displacements. Shaded lines plot the variances calculated from 50 individual bootstrap samples, and solid lines correspond to their average. We correct for preparation and measurement errors by offsetting the variances with a baseline measurement obtained without noise injection. Dashed lines for dephasing noise correspond to numerical simulations of Eq.~\ref{eq:noisy_alpha_tot}. The simulated variance under heating is that of a Binomial distribution, $\mathbb{V}[\tilde{\mathcal{F}}] = p(1-p)/M$, where $p = \mathbb{E}[\tilde{\mathcal{F}}]$ is the measurement probability.}
    \label{fig:exp_variance}
\end{figure}

\subsection{BRB validation under engineered noise}

The error models of Sec.~\ref{sec:error_model} are experimentally validated by engineering noise in the trapped ion system and performing BRB.  Here, we describe the noise-engineering protocol and present experimental measurements validating the key predictions in Sec.~\ref{sec:error_model} for the behaviour of $\mathbb{E}[\tilde{\mathcal{F}}]$ and $\mathbb{V}[\tilde{\mathcal{F}}]$ under different noise models.

Heating is controllably engineered by injecting electric field noise directly into the ion trap hardware. A signal oscillating near the ion's radial motional frequency is capacitively coupled onto one of the trap's compensation electrodes, located about $\SI{4.8}{mm}$ from the ion's position. The signal is generated by an arbitrary waveform generator (Keysight 33600a) with waveforms designed to incorporate white noise in a 40 kHz bandwidth centred around the ion's mode frequency. The heating rate is adjusted by varying the noise amplitude and is experimentally calibrated from a standard sideband thermometry experiment \cite{Monroe1995, Turchette2000}.

Dephasing noise is engineered in the waveform generation software by adding frequency offsets to the Raman beam during the sequence of displacements. A detuning of the form $\delta \hat{a}^\dagger \hat{a}$ is introduced by changing the red and blue sideband frequencies such that $\omega_b \rightarrow \omega_b + \delta$ and $\omega_r \rightarrow \omega_r - \delta$. With these changes, the SDF Hamiltonian of Eq.~\ref{eq:sdf_hamiltonian} becomes
\begin{equation}
    H_\mathrm{SDF} = \frac{\Omega}{2}\hat{\sigma}_x \hat{a} e^{- i (\phi + \delta t)} + \mathrm{h.c.},
\end{equation}
which corresponds to the control Hamiltonian, $\tilde{H}_\mathrm{c}$, of Eq.~\ref{eq:control_hamiltonian_dephasing} that models dephasing noise. The detuning $\delta \sim \mathcal{N}(0, \sigma^2_\delta)$ is introduced in software by randomly sampling from a normal distribution with variance $\sigma^2_\delta$. The correlation length of the noise is controlled by the rate at which the detunings are updated. The updates are synchronized with the BRB sequence by changing the detuning after $\mathcal{M}_n$ displacements, where $\mathcal{M}_n$ is the correlation length of the noise.

We execute the BRB protocol under each of the engineered noise described above and show measured results for a variety of different parameters in Fig~\ref{fig:exp_validation}. We plot the decay of the mean fidelity, $\mathbb{E}[\tilde{\mathcal{F}}]$, the variance, $\mathbb{V}[\tilde{\mathcal{F}}]$, and the probability distributions of $\tilde{\mathcal{F}}$ at several sequence lengths, $L$ (Fig.~\ref{fig:exp_validation}(b,e,h)). 

First, we observe that the experimental results (blue squares) under heating and dephasing processes exhibit distinctly different functional forms for the decay of $\mathbb{E}[\tilde{\mathcal{F}}]$. Under heating noise, the mean fidelity quickly decays with small $L$, and slows down with larger $L$. Conversely, under dephasing errors, the decay is slow for small $L$ and rapidly grows for large $L$. Second, we observe a distinct behaviour in the variances, $\mathbb{V}[\tilde{\mathcal{F}}]$, for heating and dephasing ( see Fig.~\ref{fig:exp_validation}(c,f,i)). Under heating errors, the variances remain small despite increasing $L$ and lower mean fidelities. 
In the presence of dephasing, however, the variance grows with increasing $L$ as the mean fidelity decays. Moreover, we notice that correlated DC noise causes a larger variance compared to uncorrelated Markovian noise for similar lengths $L$.

The experimental measurements of Fig.~\ref{fig:exp_validation} (blue squares) are in good general agreement with the theoretical models (solid lines) summarized in Table~\ref{tab:error_model}. Furthermore, the theoretical models agree with numerical simulations (dashed lines) for mean fidelities, $\mathbb{E}[\tilde{\mathcal{F}}]$, that are close to 1. Simulations numerically integrate the parasitic displacements, $\alpha_\epsilon$, and the noise-averaged fidelity is obtained from $10^{3}$ different noise realisations at each randomisation of the BRB protocol. We then repeat this over 100 circuit randomisations at each length, $L$. 

We observe two points of discrepancy between experiment and theory. First, the experimentally measured means under dephasing noise deviate from theory as $L$ increases, due to higher order terms in Eq.~\ref{eq:alpha_epsilon_dephasing} that are neglected in the derivation here (see Appendix~\ref{app:dephasing_noise_model}). A more significant discrepancy can be observed for heating noise due to measurement imperfections that accumulate with larger errors (see Appendix~\ref{app:displacements}). These effects are well captured by numerical simulations that take into account the measurement errors (dotted line). 

An examination of the experimentally measured distributions arising from each noise source reveals a characteristic form of the distribution over randomizations at fixed $L$ (Fig.~\ref{fig:exp_validation}(c,f,i)), consistent with the theoretical predictions above. First, heating noise results in an approximately symmetric distribution about the mean, while both forms of dephasing result in skewed distributions with large variance. Furthermore, in the presence of dephasing, the skewness increases with the correlation length, $\mathcal{M}_n$. 
These distributions are validated to be consistent with $\Gamma$ distributions by overlaying the calculated distribution derived from the measured $\mathbb{E}[\tilde{\mathcal{F}}]$ and $\mathbb{V}[\tilde{\mathcal{F}}]$. Data and theory show good agreement across all noise processes and values of $L$ tested.

\begin{figure}[t]
    \centering
    \includegraphics{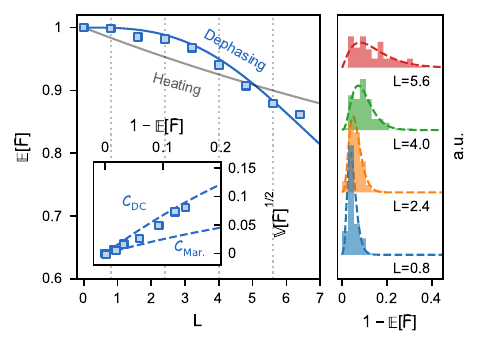}
    \caption{Experimental benchmarking with bosonic randomized benchmarking. Fidelities are measured with varying sequence lengths $L=|\alpha_0| J$, where $|\alpha_0| = 0.2$. We normalize the means and offset the variances by the measurements at $L=0$, corresponding to $\mathbb{E} = 0.96$ and $\mathbb{V} = \num{4.8e-4}$, respectively. Solid lines are fits to the dephasing (blue) and heating (grey) error models of Table~\ref{tab:error_model}. Dashed lines in the inset plot the theoretical variance under Markovian and DC noise.}
    \label{fig:exp_benchmarking}
\end{figure}

We next verify the behaviour of the variance with the number of noise averages, or measurement repetitions, for each noise mechanism (see Fig.~\ref{fig:exp_variance}). We fix $L=3.2$ and perform $M = 10^{3}$ measurement repetitions with different noise realisations and bootstrap the results to obtain variances for a range $M \in [1, 1000]$~\cite{Efron1994}. For small $M$, the variances are dominated by quantum projection noise and follow $\mathbb{V} \propto M^{-1}$. The variance under heating continues to decrease for higher noise averages, since $\mathbb{V} = 0$ in the limit of large $M$ (see Eq.~\ref{eq:mean_var_heating}). However, the variance for both correlated and uncorrelated dephasing noise begins to plateau around $M = 10$ and $M = 100$, respectively. This confirms the observations made in Section~\ref{sec:error_model} that, even in the case where dephasing noise is Markovian and uncorrelated, the variance is non-zero since the fidelity is correlated with the distance from the origin in phase space taken by the randomized trajectories.

These experiments using engineered noise validate the key theoretical predictions of this model. In the following, we apply BRB to characterize and benchmark the intrinsic noise of our system. 

\subsection{Intrinsic hardware noise characterization}
\label{sec:hardware_characterization}

We now apply the BRB protocol to benchmark a bosonic mode in the trapped ion's radial motion under intrinsic noise (see Fig.~\ref{fig:exp_benchmarking}). We plot the mean ($\mathbb{E}[\tilde{\mathcal{F}}]$), variance ($\mathbb{V}[\tilde{\mathcal{F}}]$), and the probability distributions of $\tilde{\mathcal{F}}$ to provide a qualitative comparison with the validated models of Fig.~\ref{fig:exp_validation}. First, we observe that the decay of $\mathbb{E}[\tilde{\mathcal{F}}]$ qualitatively matches a decay under dephasing.  With increasing $L$ we observe an initial slow decay which accelerates, as distinguished from the rapid decay expected from the presence of heating. Second, the variance of the distribution over randomizations is large and positively skewed (see the right-hand panel of Fig.~\ref{fig:exp_benchmarking}), suggesting that the noise is correlated.    

We quantitatively analyse the results of Fig.~\ref{fig:exp_benchmarking} by fitting the mean of the measured fidelities to the error models of Table~\ref{tab:error_model}. The most likely model is determined from the Akaike information criteria (AIC), where a smaller value indicates a better model (see Appendix~\ref{app:aic} for details). We find $\mathrm{AIC}_\mathrm{h} = -73.1$ and $\mathrm{AIC}_\mathrm{d} = -76.9$ for heating and dephasing, respectively, which confirms our previous observation that dephasing is the dominant noise mechanism. The decay rate of $\mathbb{E}[\tilde{\mathcal{F}}]$ extracted from the fit is $\eta = 0.085(2)$. Since $\eta \ll 1$, we are in a regime where the theoretical error model is in good agreement with the exact dynamics, further validating the fit results (see Appendix \ref{app:dephasing_noise_model}). We visually deduce from the variance that the noise is highly correlated quasi-static DC noise with a correlation length $M_n = J$ (see inset of Fig.~\ref{fig:exp_benchmarking}). The dephasing noise standard deviation is found from the error model of Table~\ref{tab:error_model}, resulting in $\sigma_\delta /2 \pi = \SI{114}{Hz}$. This is in good agreement with interleaved radial mode frequency calibration measurements taken throughout the experiment.

\section{Conclusion}

In summary, we have outlined a bosonic randomized benchmarking (BRB) protocol to characterize bosonic modes. The scheme consists of applying a series of randomized displacements, followed by a revival and a fidelity measurement. We provide analytical error models for three noise sources: heating, uncorrelated (Markovian) dephasing, and correlated (DC) dephasing. The combined behaviour of the mean and variance of the fidelity is distinct for each noise mechanism and correlation. Therefore, from a few measurements, one can extract the dominant noise mechanism, its error rate, and noise correlation. We experimentally validate the analytical error models on a trapped ion system by engineering each noise source and verifying the evolution of the means and variances of the fidelity. We further apply the BRB protocol to benchmark our system, and find that correlated dephasing noise is the dominant error mechanism. 

In Ref.~\cite{Ball2016}, it was found that the means and variances of the fidelity under Randomized Benchmarking of qubits can be used to retrieve the power spectral density of the noise. A similar relation could be derived for the dephasing noise considered here, and we reserve this for future work. Looking forward, our BRB scheme can be extended to study error correlations between different bosonic modes. One can also consider different error mechanisms, such as separate photon gains or losses.

\section{Acknowledgements}

We were supported by the U.S. Office of Naval Research Global (N62909-20-1-2047), the U.S. Army Research Office Laboratory for Physical Sciences (W911NF-21-1-0003), the U.S. Air Force Office of Scientific Research (FA2386-23-1-4062), the U.S. Intelligence Advanced Research Projects Activity (W911NF-16-1-0070), Lockheed Martin, the Sydney Quantum Academy (TRT), the Wellcome Leap Quantum for Bio program, the Australian Research Council, and H. and A. Harley.

\section{Data availability}

The experimental data are available upon reasonable request.

\bibliography{bib.bib}

\begin{thebibliography}{68}%
\makeatletter
\providecommand \@ifxundefined [1]{%
 \@ifx{#1\undefined}
}%
\providecommand \@ifnum [1]{%
 \ifnum #1\expandafter \@firstoftwo
 \else \expandafter \@secondoftwo
 \fi
}%
\providecommand \@ifx [1]{%
 \ifx #1\expandafter \@firstoftwo
 \else \expandafter \@secondoftwo
 \fi
}%
\providecommand \natexlab [1]{#1}%
\providecommand \enquote  [1]{``#1''}%
\providecommand \bibnamefont  [1]{#1}%
\providecommand \bibfnamefont [1]{#1}%
\providecommand \citenamefont [1]{#1}%
\providecommand \href@noop [0]{\@secondoftwo}%
\providecommand \href [0]{\begingroup \@sanitize@url \@href}%
\providecommand \@href[1]{\@@startlink{#1}\@@href}%
\providecommand \@@href[1]{\endgroup#1\@@endlink}%
\providecommand \@sanitize@url [0]{\catcode `\\12\catcode `\$12\catcode
  `\&12\catcode `\#12\catcode `\^12\catcode `\_12\catcode `\%12\relax}%
\providecommand \@@startlink[1]{}%
\providecommand \@@endlink[0]{}%
\providecommand \url  [0]{\begingroup\@sanitize@url \@url }%
\providecommand \@url [1]{\endgroup\@href {#1}{\urlprefix }}%
\providecommand \urlprefix  [0]{URL }%
\providecommand \Eprint [0]{\href }%
\providecommand \doibase [0]{http://dx.doi.org/}%
\providecommand \selectlanguage [0]{\@gobble}%
\providecommand \bibinfo  [0]{\@secondoftwo}%
\providecommand \bibfield  [0]{\@secondoftwo}%
\providecommand \translation [1]{[#1]}%
\providecommand \BibitemOpen [0]{}%
\providecommand \bibitemStop [0]{}%
\providecommand \bibitemNoStop [0]{.\EOS\space}%
\providecommand \EOS [0]{\spacefactor3000\relax}%
\providecommand \BibitemShut  [1]{\csname bibitem#1\endcsname}%
\let\auto@bib@innerbib\@empty
\bibitem [{\citenamefont {Sørensen}\ and\ \citenamefont
  {Mølmer}(1999)}]{Srensen1999}%
  \BibitemOpen
  \bibfield  {author} {\bibinfo {author} {\bibfnamefont {A.}~\bibnamefont
  {Sørensen}}\ and\ \bibinfo {author} {\bibfnamefont {K.}~\bibnamefont
  {Mølmer}},\ }\href {\doibase 10.1103/physrevlett.82.1971} {\bibfield
  {journal} {\bibinfo  {journal} {Physical Review Letters}\ }\textbf {\bibinfo
  {volume} {82}},\ \bibinfo {pages} {1971–1974} (\bibinfo {year}
  {1999})}\BibitemShut {NoStop}%
\bibitem [{\citenamefont {Knill}\ \emph {et~al.}(2001)\citenamefont {Knill},
  \citenamefont {Laflamme},\ and\ \citenamefont {Milburn}}]{Knill2001}%
  \BibitemOpen
  \bibfield  {author} {\bibinfo {author} {\bibfnamefont {E.}~\bibnamefont
  {Knill}}, \bibinfo {author} {\bibfnamefont {R.}~\bibnamefont {Laflamme}}, \
  and\ \bibinfo {author} {\bibfnamefont {G.~J.}\ \bibnamefont {Milburn}},\
  }\href {\doibase 10.1038/35051009} {\bibfield  {journal} {\bibinfo  {journal}
  {Nature}\ }\textbf {\bibinfo {volume} {409}},\ \bibinfo {pages} {46–52}
  (\bibinfo {year} {2001})}\BibitemShut {NoStop}%
\bibitem [{\citenamefont {Braunstein}\ and\ \citenamefont {van
  Loock}(2005)}]{Braunstein2005}%
  \BibitemOpen
  \bibfield  {author} {\bibinfo {author} {\bibfnamefont {S.~L.}\ \bibnamefont
  {Braunstein}}\ and\ \bibinfo {author} {\bibfnamefont {P.}~\bibnamefont {van
  Loock}},\ }\href {\doibase 10.1103/revmodphys.77.513} {\bibfield  {journal}
  {\bibinfo  {journal} {Reviews of Modern Physics}\ }\textbf {\bibinfo {volume}
  {77}},\ \bibinfo {pages} {513–577} (\bibinfo {year} {2005})}\BibitemShut
  {NoStop}%
\bibitem [{\citenamefont {Weedbrook}\ \emph {et~al.}(2012)\citenamefont
  {Weedbrook}, \citenamefont {Pirandola}, \citenamefont {García-Patrón},
  \citenamefont {Cerf}, \citenamefont {Ralph}, \citenamefont {Shapiro},\ and\
  \citenamefont {Lloyd}}]{Weedbrook2012}%
  \BibitemOpen
  \bibfield  {author} {\bibinfo {author} {\bibfnamefont {C.}~\bibnamefont
  {Weedbrook}}, \bibinfo {author} {\bibfnamefont {S.}~\bibnamefont
  {Pirandola}}, \bibinfo {author} {\bibfnamefont {R.}~\bibnamefont
  {García-Patrón}}, \bibinfo {author} {\bibfnamefont {N.~J.}\ \bibnamefont
  {Cerf}}, \bibinfo {author} {\bibfnamefont {T.~C.}\ \bibnamefont {Ralph}},
  \bibinfo {author} {\bibfnamefont {J.~H.}\ \bibnamefont {Shapiro}}, \ and\
  \bibinfo {author} {\bibfnamefont {S.}~\bibnamefont {Lloyd}},\ }\href
  {\doibase 10.1103/revmodphys.84.621} {\bibfield  {journal} {\bibinfo
  {journal} {Reviews of Modern Physics}\ }\textbf {\bibinfo {volume} {84}},\
  \bibinfo {pages} {621–669} (\bibinfo {year} {2012})}\BibitemShut {NoStop}%
\bibitem [{\citenamefont {Giovannetti}\ \emph {et~al.}(2004)\citenamefont
  {Giovannetti}, \citenamefont {Lloyd},\ and\ \citenamefont
  {Maccone}}]{Giovannetti2004}%
  \BibitemOpen
  \bibfield  {author} {\bibinfo {author} {\bibfnamefont {V.}~\bibnamefont
  {Giovannetti}}, \bibinfo {author} {\bibfnamefont {S.}~\bibnamefont {Lloyd}},
  \ and\ \bibinfo {author} {\bibfnamefont {L.}~\bibnamefont {Maccone}},\ }\href
  {\doibase 10.1126/science.1104149} {\bibfield  {journal} {\bibinfo  {journal}
  {Science}\ }\textbf {\bibinfo {volume} {306}},\ \bibinfo {pages}
  {1330–1336} (\bibinfo {year} {2004})}\BibitemShut {NoStop}%
\bibitem [{\citenamefont {Gilmore}\ \emph {et~al.}(2021)\citenamefont
  {Gilmore}, \citenamefont {Affolter}, \citenamefont {Lewis-Swan},
  \citenamefont {Barberena}, \citenamefont {Jordan}, \citenamefont {Rey},\ and\
  \citenamefont {Bollinger}}]{Gilmore2021}%
  \BibitemOpen
  \bibfield  {author} {\bibinfo {author} {\bibfnamefont {K.~A.}\ \bibnamefont
  {Gilmore}}, \bibinfo {author} {\bibfnamefont {M.}~\bibnamefont {Affolter}},
  \bibinfo {author} {\bibfnamefont {R.~J.}\ \bibnamefont {Lewis-Swan}},
  \bibinfo {author} {\bibfnamefont {D.}~\bibnamefont {Barberena}}, \bibinfo
  {author} {\bibfnamefont {E.}~\bibnamefont {Jordan}}, \bibinfo {author}
  {\bibfnamefont {A.~M.}\ \bibnamefont {Rey}}, \ and\ \bibinfo {author}
  {\bibfnamefont {J.~J.}\ \bibnamefont {Bollinger}},\ }\href {\doibase
  10.1126/science.abi5226} {\bibfield  {journal} {\bibinfo  {journal}
  {Science}\ }\textbf {\bibinfo {volume} {373}},\ \bibinfo {pages} {673}
  (\bibinfo {year} {2021})}\BibitemShut {NoStop}%
\bibitem [{\citenamefont {Bruzewicz}\ \emph {et~al.}(2019)\citenamefont
  {Bruzewicz}, \citenamefont {Chiaverini}, \citenamefont {McConnell},\ and\
  \citenamefont {Sage}}]{Bruzewicz2019}%
  \BibitemOpen
  \bibfield  {author} {\bibinfo {author} {\bibfnamefont {C.~D.}\ \bibnamefont
  {Bruzewicz}}, \bibinfo {author} {\bibfnamefont {J.}~\bibnamefont
  {Chiaverini}}, \bibinfo {author} {\bibfnamefont {R.}~\bibnamefont
  {McConnell}}, \ and\ \bibinfo {author} {\bibfnamefont {J.~M.}\ \bibnamefont
  {Sage}},\ }\href {\doibase 10.1063/1.5088164} {\bibfield  {journal} {\bibinfo
   {journal} {Applied Physics Reviews}\ }\textbf {\bibinfo {volume} {6}}
  (\bibinfo {year} {2019}),\ 10.1063/1.5088164}\BibitemShut {NoStop}%
\bibitem [{\citenamefont {Blais}\ \emph {et~al.}(2021)\citenamefont {Blais},
  \citenamefont {Grimsmo}, \citenamefont {Girvin},\ and\ \citenamefont
  {Wallraff}}]{Blais2021}%
  \BibitemOpen
  \bibfield  {author} {\bibinfo {author} {\bibfnamefont {A.}~\bibnamefont
  {Blais}}, \bibinfo {author} {\bibfnamefont {A.~L.}\ \bibnamefont {Grimsmo}},
  \bibinfo {author} {\bibfnamefont {S.}~\bibnamefont {Girvin}}, \ and\ \bibinfo
  {author} {\bibfnamefont {A.}~\bibnamefont {Wallraff}},\ }\href {\doibase
  10.1103/revmodphys.93.025005} {\bibfield  {journal} {\bibinfo  {journal}
  {Reviews of Modern Physics}\ }\textbf {\bibinfo {volume} {93}} (\bibinfo
  {year} {2021}),\ 10.1103/revmodphys.93.025005}\BibitemShut {NoStop}%
\bibitem [{\citenamefont {Bourassa}\ \emph {et~al.}(2021)\citenamefont
  {Bourassa}, \citenamefont {Alexander}, \citenamefont {Vasmer}, \citenamefont
  {Patil}, \citenamefont {Tzitrin}, \citenamefont {Matsuura}, \citenamefont
  {Su}, \citenamefont {Baragiola}, \citenamefont {Guha}, \citenamefont
  {Dauphinais}, \citenamefont {Sabapathy}, \citenamefont {Menicucci},\ and\
  \citenamefont {Dhand}}]{Bourassa2021}%
  \BibitemOpen
  \bibfield  {author} {\bibinfo {author} {\bibfnamefont {J.~E.}\ \bibnamefont
  {Bourassa}}, \bibinfo {author} {\bibfnamefont {R.~N.}\ \bibnamefont
  {Alexander}}, \bibinfo {author} {\bibfnamefont {M.}~\bibnamefont {Vasmer}},
  \bibinfo {author} {\bibfnamefont {A.}~\bibnamefont {Patil}}, \bibinfo
  {author} {\bibfnamefont {I.}~\bibnamefont {Tzitrin}}, \bibinfo {author}
  {\bibfnamefont {T.}~\bibnamefont {Matsuura}}, \bibinfo {author}
  {\bibfnamefont {D.}~\bibnamefont {Su}}, \bibinfo {author} {\bibfnamefont
  {B.~Q.}\ \bibnamefont {Baragiola}}, \bibinfo {author} {\bibfnamefont
  {S.}~\bibnamefont {Guha}}, \bibinfo {author} {\bibfnamefont {G.}~\bibnamefont
  {Dauphinais}}, \bibinfo {author} {\bibfnamefont {K.~K.}\ \bibnamefont
  {Sabapathy}}, \bibinfo {author} {\bibfnamefont {N.~C.}\ \bibnamefont
  {Menicucci}}, \ and\ \bibinfo {author} {\bibfnamefont {I.}~\bibnamefont
  {Dhand}},\ }\href {\doibase 10.22331/q-2021-02-04-392} {\bibfield  {journal}
  {\bibinfo  {journal} {Quantum}\ }\textbf {\bibinfo {volume} {5}},\ \bibinfo
  {pages} {392} (\bibinfo {year} {2021})}\BibitemShut {NoStop}%
\bibitem [{\citenamefont {Fl\"{u}hmann}\ \emph {et~al.}(2019)\citenamefont
  {Fl\"{u}hmann}, \citenamefont {Nguyen}, \citenamefont {Marinelli},
  \citenamefont {Negnevitsky}, \citenamefont {Mehta},\ and\ \citenamefont
  {Home}}]{Fluhmann2019}%
  \BibitemOpen
  \bibfield  {author} {\bibinfo {author} {\bibfnamefont {C.}~\bibnamefont
  {Fl\"{u}hmann}}, \bibinfo {author} {\bibfnamefont {T.~L.}\ \bibnamefont
  {Nguyen}}, \bibinfo {author} {\bibfnamefont {M.}~\bibnamefont {Marinelli}},
  \bibinfo {author} {\bibfnamefont {V.}~\bibnamefont {Negnevitsky}}, \bibinfo
  {author} {\bibfnamefont {K.}~\bibnamefont {Mehta}}, \ and\ \bibinfo {author}
  {\bibfnamefont {J.~P.}\ \bibnamefont {Home}},\ }\href {\doibase
  10.1038/s41586-019-0960-6} {\bibfield  {journal} {\bibinfo  {journal}
  {Nature}\ }\textbf {\bibinfo {volume} {566}},\ \bibinfo {pages} {513}
  (\bibinfo {year} {2019})}\BibitemShut {NoStop}%
\bibitem [{\citenamefont {Campagne-Ibarcq}\ \emph {et~al.}(2020)\citenamefont
  {Campagne-Ibarcq}, \citenamefont {Eickbusch}, \citenamefont {Touzard},
  \citenamefont {Zalys-Geller}, \citenamefont {Frattini}, \citenamefont
  {Sivak}, \citenamefont {Reinhold}, \citenamefont {Puri}, \citenamefont
  {Shankar}, \citenamefont {Schoelkopf}, \citenamefont {Frunzio}, \citenamefont
  {Mirrahimi},\ and\ \citenamefont {Devoret}}]{CampagneIbarcq2020}%
  \BibitemOpen
  \bibfield  {author} {\bibinfo {author} {\bibfnamefont {P.}~\bibnamefont
  {Campagne-Ibarcq}}, \bibinfo {author} {\bibfnamefont {A.}~\bibnamefont
  {Eickbusch}}, \bibinfo {author} {\bibfnamefont {S.}~\bibnamefont {Touzard}},
  \bibinfo {author} {\bibfnamefont {E.}~\bibnamefont {Zalys-Geller}}, \bibinfo
  {author} {\bibfnamefont {N.~E.}\ \bibnamefont {Frattini}}, \bibinfo {author}
  {\bibfnamefont {V.~V.}\ \bibnamefont {Sivak}}, \bibinfo {author}
  {\bibfnamefont {P.}~\bibnamefont {Reinhold}}, \bibinfo {author}
  {\bibfnamefont {S.}~\bibnamefont {Puri}}, \bibinfo {author} {\bibfnamefont
  {S.}~\bibnamefont {Shankar}}, \bibinfo {author} {\bibfnamefont {R.~J.}\
  \bibnamefont {Schoelkopf}}, \bibinfo {author} {\bibfnamefont
  {L.}~\bibnamefont {Frunzio}}, \bibinfo {author} {\bibfnamefont
  {M.}~\bibnamefont {Mirrahimi}}, \ and\ \bibinfo {author} {\bibfnamefont
  {M.~H.}\ \bibnamefont {Devoret}},\ }\href {\doibase
  10.1038/s41586-020-2603-3} {\bibfield  {journal} {\bibinfo  {journal}
  {Nature}\ }\textbf {\bibinfo {volume} {584}},\ \bibinfo {pages} {368}
  (\bibinfo {year} {2020})}\BibitemShut {NoStop}%
\bibitem [{\citenamefont {de~Neeve}\ \emph {et~al.}(2022)\citenamefont
  {de~Neeve}, \citenamefont {Nguyen}, \citenamefont {Behrle},\ and\
  \citenamefont {Home}}]{deNeeve2022}%
  \BibitemOpen
  \bibfield  {author} {\bibinfo {author} {\bibfnamefont {B.}~\bibnamefont
  {de~Neeve}}, \bibinfo {author} {\bibfnamefont {T.-L.}\ \bibnamefont
  {Nguyen}}, \bibinfo {author} {\bibfnamefont {T.}~\bibnamefont {Behrle}}, \
  and\ \bibinfo {author} {\bibfnamefont {J.~P.}\ \bibnamefont {Home}},\ }\href
  {\doibase 10.1038/s41567-021-01487-7} {\bibfield  {journal} {\bibinfo
  {journal} {Nature Physics}\ }\textbf {\bibinfo {volume} {18}},\ \bibinfo
  {pages} {296} (\bibinfo {year} {2022})}\BibitemShut {NoStop}%
\bibitem [{\citenamefont {Matsos}\ \emph {et~al.}(2023)\citenamefont {Matsos},
  \citenamefont {Valahu}, \citenamefont {Navickas}, \citenamefont {Rao},
  \citenamefont {Millican}, \citenamefont {Biercuk},\ and\ \citenamefont
  {Tan}}]{Matsos2023}%
  \BibitemOpen
  \bibfield  {author} {\bibinfo {author} {\bibfnamefont {V.~G.}\ \bibnamefont
  {Matsos}}, \bibinfo {author} {\bibfnamefont {C.~H.}\ \bibnamefont {Valahu}},
  \bibinfo {author} {\bibfnamefont {T.}~\bibnamefont {Navickas}}, \bibinfo
  {author} {\bibfnamefont {A.~D.}\ \bibnamefont {Rao}}, \bibinfo {author}
  {\bibfnamefont {M.~J.}\ \bibnamefont {Millican}}, \bibinfo {author}
  {\bibfnamefont {M.~J.}\ \bibnamefont {Biercuk}}, \ and\ \bibinfo {author}
  {\bibfnamefont {T.~R.}\ \bibnamefont {Tan}},\ }\href {\doibase
  10.48550/ARXIV.2310.15546} {\enquote {\bibinfo {title} {Robust and
  deterministic preparation of bosonic logical states in a trapped ion},}\ }
  (\bibinfo {year} {2023})\BibitemShut {NoStop}%
\bibitem [{\citenamefont {Heeres}\ \emph {et~al.}(2017)\citenamefont {Heeres},
  \citenamefont {Reinhold}, \citenamefont {Ofek}, \citenamefont {Frunzio},
  \citenamefont {Jiang}, \citenamefont {Devoret},\ and\ \citenamefont
  {Schoelkopf}}]{Heeres2017}%
  \BibitemOpen
  \bibfield  {author} {\bibinfo {author} {\bibfnamefont {R.~W.}\ \bibnamefont
  {Heeres}}, \bibinfo {author} {\bibfnamefont {P.}~\bibnamefont {Reinhold}},
  \bibinfo {author} {\bibfnamefont {N.}~\bibnamefont {Ofek}}, \bibinfo {author}
  {\bibfnamefont {L.}~\bibnamefont {Frunzio}}, \bibinfo {author} {\bibfnamefont
  {L.}~\bibnamefont {Jiang}}, \bibinfo {author} {\bibfnamefont {M.~H.}\
  \bibnamefont {Devoret}}, \ and\ \bibinfo {author} {\bibfnamefont {R.~J.}\
  \bibnamefont {Schoelkopf}},\ }\href {\doibase 10.1038/s41467-017-00045-1}
  {\bibfield  {journal} {\bibinfo  {journal} {Nature Communications}\ }\textbf
  {\bibinfo {volume} {8}} (\bibinfo {year} {2017}),\
  10.1038/s41467-017-00045-1}\BibitemShut {NoStop}%
\bibitem [{\citenamefont {Gertler}\ \emph {et~al.}(2021)\citenamefont
  {Gertler}, \citenamefont {Baker}, \citenamefont {Li}, \citenamefont {Shirol},
  \citenamefont {Koch},\ and\ \citenamefont {Wang}}]{Gertler2021}%
  \BibitemOpen
  \bibfield  {author} {\bibinfo {author} {\bibfnamefont {J.~M.}\ \bibnamefont
  {Gertler}}, \bibinfo {author} {\bibfnamefont {B.}~\bibnamefont {Baker}},
  \bibinfo {author} {\bibfnamefont {J.}~\bibnamefont {Li}}, \bibinfo {author}
  {\bibfnamefont {S.}~\bibnamefont {Shirol}}, \bibinfo {author} {\bibfnamefont
  {J.}~\bibnamefont {Koch}}, \ and\ \bibinfo {author} {\bibfnamefont
  {C.}~\bibnamefont {Wang}},\ }\href {\doibase 10.1038/s41586-021-03257-0}
  {\bibfield  {journal} {\bibinfo  {journal} {Nature}\ }\textbf {\bibinfo
  {volume} {590}},\ \bibinfo {pages} {243} (\bibinfo {year}
  {2021})}\BibitemShut {NoStop}%
\bibitem [{\citenamefont {Lachance-Quirion}\ \emph {et~al.}(2023)\citenamefont
  {Lachance-Quirion}, \citenamefont {Lemonde}, \citenamefont {Simoneau},
  \citenamefont {St-Jean}, \citenamefont {Lemieux}, \citenamefont {Turcotte},
  \citenamefont {Wright}, \citenamefont {Lacroix}, \citenamefont
  {Fréchette-Viens}, \citenamefont {Shillito}, \citenamefont {Hopfmueller},
  \citenamefont {Tremblay}, \citenamefont {Frattini}, \citenamefont {Lemyre},\
  and\ \citenamefont {St-Jean}}]{lachancequirion2023autonomous}%
  \BibitemOpen
  \bibfield  {author} {\bibinfo {author} {\bibfnamefont {D.}~\bibnamefont
  {Lachance-Quirion}}, \bibinfo {author} {\bibfnamefont {M.-A.}\ \bibnamefont
  {Lemonde}}, \bibinfo {author} {\bibfnamefont {J.~O.}\ \bibnamefont
  {Simoneau}}, \bibinfo {author} {\bibfnamefont {L.}~\bibnamefont {St-Jean}},
  \bibinfo {author} {\bibfnamefont {P.}~\bibnamefont {Lemieux}}, \bibinfo
  {author} {\bibfnamefont {S.}~\bibnamefont {Turcotte}}, \bibinfo {author}
  {\bibfnamefont {W.}~\bibnamefont {Wright}}, \bibinfo {author} {\bibfnamefont
  {A.}~\bibnamefont {Lacroix}}, \bibinfo {author} {\bibfnamefont
  {J.}~\bibnamefont {Fréchette-Viens}}, \bibinfo {author} {\bibfnamefont
  {R.}~\bibnamefont {Shillito}}, \bibinfo {author} {\bibfnamefont
  {F.}~\bibnamefont {Hopfmueller}}, \bibinfo {author} {\bibfnamefont
  {M.}~\bibnamefont {Tremblay}}, \bibinfo {author} {\bibfnamefont {N.~E.}\
  \bibnamefont {Frattini}}, \bibinfo {author} {\bibfnamefont {J.~C.}\
  \bibnamefont {Lemyre}}, \ and\ \bibinfo {author} {\bibfnamefont
  {P.}~\bibnamefont {St-Jean}},\ }\href@noop {} {\enquote {\bibinfo {title}
  {Autonomous quantum error correction of gottesman-kitaev-preskill states},}\
  } (\bibinfo {year} {2023}),\ \Eprint {http://arxiv.org/abs/2310.11400}
  {arXiv:2310.11400 [quant-ph]} \BibitemShut {NoStop}%
\bibitem [{\citenamefont {Hu}\ \emph {et~al.}(2019)\citenamefont {Hu},
  \citenamefont {Ma}, \citenamefont {Cai}, \citenamefont {Mu}, \citenamefont
  {Xu}, \citenamefont {Wang}, \citenamefont {Wu}, \citenamefont {Wang},
  \citenamefont {Song}, \citenamefont {Zou}, \citenamefont {Girvin},
  \citenamefont {Duan},\ and\ \citenamefont {Sun}}]{Hu2019}%
  \BibitemOpen
  \bibfield  {author} {\bibinfo {author} {\bibfnamefont {L.}~\bibnamefont
  {Hu}}, \bibinfo {author} {\bibfnamefont {Y.}~\bibnamefont {Ma}}, \bibinfo
  {author} {\bibfnamefont {W.}~\bibnamefont {Cai}}, \bibinfo {author}
  {\bibfnamefont {X.}~\bibnamefont {Mu}}, \bibinfo {author} {\bibfnamefont
  {Y.}~\bibnamefont {Xu}}, \bibinfo {author} {\bibfnamefont {W.}~\bibnamefont
  {Wang}}, \bibinfo {author} {\bibfnamefont {Y.}~\bibnamefont {Wu}}, \bibinfo
  {author} {\bibfnamefont {H.}~\bibnamefont {Wang}}, \bibinfo {author}
  {\bibfnamefont {Y.~P.}\ \bibnamefont {Song}}, \bibinfo {author}
  {\bibfnamefont {C.-L.}\ \bibnamefont {Zou}}, \bibinfo {author} {\bibfnamefont
  {S.~M.}\ \bibnamefont {Girvin}}, \bibinfo {author} {\bibfnamefont {L.-M.}\
  \bibnamefont {Duan}}, \ and\ \bibinfo {author} {\bibfnamefont
  {L.}~\bibnamefont {Sun}},\ }\href {\doibase 10.1038/s41567-018-0414-3}
  {\bibfield  {journal} {\bibinfo  {journal} {Nature Physics}\ }\textbf
  {\bibinfo {volume} {15}},\ \bibinfo {pages} {503} (\bibinfo {year}
  {2019})}\BibitemShut {NoStop}%
\bibitem [{\citenamefont {Eickbusch}\ \emph {et~al.}(2022)\citenamefont
  {Eickbusch}, \citenamefont {Sivak}, \citenamefont {Ding}, \citenamefont
  {Elder}, \citenamefont {Jha}, \citenamefont {Venkatraman}, \citenamefont
  {Royer}, \citenamefont {Girvin}, \citenamefont {Schoelkopf},\ and\
  \citenamefont {Devoret}}]{Eickbusch2022}%
  \BibitemOpen
  \bibfield  {author} {\bibinfo {author} {\bibfnamefont {A.}~\bibnamefont
  {Eickbusch}}, \bibinfo {author} {\bibfnamefont {V.}~\bibnamefont {Sivak}},
  \bibinfo {author} {\bibfnamefont {A.~Z.}\ \bibnamefont {Ding}}, \bibinfo
  {author} {\bibfnamefont {S.~S.}\ \bibnamefont {Elder}}, \bibinfo {author}
  {\bibfnamefont {S.~R.}\ \bibnamefont {Jha}}, \bibinfo {author} {\bibfnamefont
  {J.}~\bibnamefont {Venkatraman}}, \bibinfo {author} {\bibfnamefont
  {B.}~\bibnamefont {Royer}}, \bibinfo {author} {\bibfnamefont {S.~M.}\
  \bibnamefont {Girvin}}, \bibinfo {author} {\bibfnamefont {R.~J.}\
  \bibnamefont {Schoelkopf}}, \ and\ \bibinfo {author} {\bibfnamefont {M.~H.}\
  \bibnamefont {Devoret}},\ }\href {\doibase 10.1038/s41567-022-01776-9}
  {\bibfield  {journal} {\bibinfo  {journal} {Nature Physics}\ }\textbf
  {\bibinfo {volume} {18}},\ \bibinfo {pages} {1464} (\bibinfo {year}
  {2022})}\BibitemShut {NoStop}%
\bibitem [{\citenamefont {Ni}\ \emph {et~al.}(2023)\citenamefont {Ni},
  \citenamefont {Li}, \citenamefont {Deng}, \citenamefont {Cai}, \citenamefont
  {Zhang}, \citenamefont {Wang}, \citenamefont {Yang}, \citenamefont {Yu},
  \citenamefont {Yan}, \citenamefont {Liu}, \citenamefont {Zou}, \citenamefont
  {Sun}, \citenamefont {Zheng}, \citenamefont {Xu},\ and\ \citenamefont
  {Yu}}]{Ni2023}%
  \BibitemOpen
  \bibfield  {author} {\bibinfo {author} {\bibfnamefont {Z.}~\bibnamefont
  {Ni}}, \bibinfo {author} {\bibfnamefont {S.}~\bibnamefont {Li}}, \bibinfo
  {author} {\bibfnamefont {X.}~\bibnamefont {Deng}}, \bibinfo {author}
  {\bibfnamefont {Y.}~\bibnamefont {Cai}}, \bibinfo {author} {\bibfnamefont
  {L.}~\bibnamefont {Zhang}}, \bibinfo {author} {\bibfnamefont
  {W.}~\bibnamefont {Wang}}, \bibinfo {author} {\bibfnamefont {Z.-B.}\
  \bibnamefont {Yang}}, \bibinfo {author} {\bibfnamefont {H.}~\bibnamefont
  {Yu}}, \bibinfo {author} {\bibfnamefont {F.}~\bibnamefont {Yan}}, \bibinfo
  {author} {\bibfnamefont {S.}~\bibnamefont {Liu}}, \bibinfo {author}
  {\bibfnamefont {C.-L.}\ \bibnamefont {Zou}}, \bibinfo {author} {\bibfnamefont
  {L.}~\bibnamefont {Sun}}, \bibinfo {author} {\bibfnamefont {S.-B.}\
  \bibnamefont {Zheng}}, \bibinfo {author} {\bibfnamefont {Y.}~\bibnamefont
  {Xu}}, \ and\ \bibinfo {author} {\bibfnamefont {D.}~\bibnamefont {Yu}},\
  }\href {\doibase 10.1038/s41586-023-05784-4} {\bibfield  {journal} {\bibinfo
  {journal} {Nature}\ }\textbf {\bibinfo {volume} {616}},\ \bibinfo {pages}
  {56} (\bibinfo {year} {2023})}\BibitemShut {NoStop}%
\bibitem [{\citenamefont {Cirac}\ and\ \citenamefont
  {Zoller}(1995)}]{Cirac1995}%
  \BibitemOpen
  \bibfield  {author} {\bibinfo {author} {\bibfnamefont {J.~I.}\ \bibnamefont
  {Cirac}}\ and\ \bibinfo {author} {\bibfnamefont {P.}~\bibnamefont {Zoller}},\
  }\href {\doibase 10.1103/physrevlett.74.4091} {\bibfield  {journal} {\bibinfo
   {journal} {Physical Review Letters}\ }\textbf {\bibinfo {volume} {74}},\
  \bibinfo {pages} {4091} (\bibinfo {year} {1995})}\BibitemShut {NoStop}%
\bibitem [{\citenamefont {S{\o}rensen}\ and\ \citenamefont
  {M{\o}lmer}(2000)}]{Srensen2000}%
  \BibitemOpen
  \bibfield  {author} {\bibinfo {author} {\bibfnamefont {A.}~\bibnamefont
  {S{\o}rensen}}\ and\ \bibinfo {author} {\bibfnamefont {K.}~\bibnamefont
  {M{\o}lmer}},\ }\href {\doibase 10.1103/physreva.62.022311} {\bibfield
  {journal} {\bibinfo  {journal} {Physical Review A}\ }\textbf {\bibinfo
  {volume} {62}} (\bibinfo {year} {2000}),\
  10.1103/physreva.62.022311}\BibitemShut {NoStop}%
\bibitem [{\citenamefont {Britton}\ \emph {et~al.}(2012)\citenamefont
  {Britton}, \citenamefont {Sawyer}, \citenamefont {Keith}, \citenamefont
  {Wang}, \citenamefont {Freericks}, \citenamefont {Uys}, \citenamefont
  {Biercuk},\ and\ \citenamefont {Bollinger}}]{britton2009}%
  \BibitemOpen
  \bibfield  {author} {\bibinfo {author} {\bibfnamefont {J.~W.}\ \bibnamefont
  {Britton}}, \bibinfo {author} {\bibfnamefont {B.~C.}\ \bibnamefont {Sawyer}},
  \bibinfo {author} {\bibfnamefont {A.~C.}\ \bibnamefont {Keith}}, \bibinfo
  {author} {\bibfnamefont {C.~C.~J.}\ \bibnamefont {Wang}}, \bibinfo {author}
  {\bibfnamefont {J.~K.}\ \bibnamefont {Freericks}}, \bibinfo {author}
  {\bibfnamefont {H.}~\bibnamefont {Uys}}, \bibinfo {author} {\bibfnamefont
  {M.~J.}\ \bibnamefont {Biercuk}}, \ and\ \bibinfo {author} {\bibfnamefont
  {J.~J.}\ \bibnamefont {Bollinger}},\ }\href {\doibase 10.1038/nature10981}
  {\bibfield  {journal} {\bibinfo  {journal} {Nature}\ }\textbf {\bibinfo
  {volume} {484}},\ \bibinfo {pages} {489} (\bibinfo {year}
  {2012})}\BibitemShut {NoStop}%
\bibitem [{\citenamefont {Leung}\ \emph {et~al.}(2018)\citenamefont {Leung},
  \citenamefont {Landsman}, \citenamefont {Figgatt}, \citenamefont {Linke},
  \citenamefont {Monroe},\ and\ \citenamefont {Brown}}]{Leung2018}%
  \BibitemOpen
  \bibfield  {author} {\bibinfo {author} {\bibfnamefont {P.~H.}\ \bibnamefont
  {Leung}}, \bibinfo {author} {\bibfnamefont {K.~A.}\ \bibnamefont {Landsman}},
  \bibinfo {author} {\bibfnamefont {C.}~\bibnamefont {Figgatt}}, \bibinfo
  {author} {\bibfnamefont {N.~M.}\ \bibnamefont {Linke}}, \bibinfo {author}
  {\bibfnamefont {C.}~\bibnamefont {Monroe}}, \ and\ \bibinfo {author}
  {\bibfnamefont {K.~R.}\ \bibnamefont {Brown}},\ }\href {\doibase
  10.1103/physrevlett.120.020501} {\bibfield  {journal} {\bibinfo  {journal}
  {Physical Review Letters}\ }\textbf {\bibinfo {volume} {120}} (\bibinfo
  {year} {2018}),\ 10.1103/physrevlett.120.020501}\BibitemShut {NoStop}%
\bibitem [{\citenamefont {Milne}\ \emph {et~al.}(2020)\citenamefont {Milne},
  \citenamefont {Edmunds}, \citenamefont {Hempel}, \citenamefont {Roy},
  \citenamefont {Mavadia},\ and\ \citenamefont {Biercuk}}]{Milne2020}%
  \BibitemOpen
  \bibfield  {author} {\bibinfo {author} {\bibfnamefont {A.~R.}\ \bibnamefont
  {Milne}}, \bibinfo {author} {\bibfnamefont {C.~L.}\ \bibnamefont {Edmunds}},
  \bibinfo {author} {\bibfnamefont {C.}~\bibnamefont {Hempel}}, \bibinfo
  {author} {\bibfnamefont {F.}~\bibnamefont {Roy}}, \bibinfo {author}
  {\bibfnamefont {S.}~\bibnamefont {Mavadia}}, \ and\ \bibinfo {author}
  {\bibfnamefont {M.~J.}\ \bibnamefont {Biercuk}},\ }\href {\doibase
  10.1103/physrevapplied.13.024022} {\bibfield  {journal} {\bibinfo  {journal}
  {Physical Review Applied}\ }\textbf {\bibinfo {volume} {13}} (\bibinfo {year}
  {2020}),\ 10.1103/physrevapplied.13.024022}\BibitemShut {NoStop}%
\bibitem [{\citenamefont {Bentley}\ \emph {et~al.}(2020)\citenamefont
  {Bentley}, \citenamefont {Ball}, \citenamefont {Biercuk}, \citenamefont
  {Carvalho}, \citenamefont {Hush},\ and\ \citenamefont
  {Slatyer}}]{Bentley2020}%
  \BibitemOpen
  \bibfield  {author} {\bibinfo {author} {\bibfnamefont {C.~D.~B.}\
  \bibnamefont {Bentley}}, \bibinfo {author} {\bibfnamefont {H.}~\bibnamefont
  {Ball}}, \bibinfo {author} {\bibfnamefont {M.~J.}\ \bibnamefont {Biercuk}},
  \bibinfo {author} {\bibfnamefont {A.~R.~R.}\ \bibnamefont {Carvalho}},
  \bibinfo {author} {\bibfnamefont {M.~R.}\ \bibnamefont {Hush}}, \ and\
  \bibinfo {author} {\bibfnamefont {H.~J.}\ \bibnamefont {Slatyer}},\ }\href
  {\doibase 10.1002/qute.202070113} {\bibfield  {journal} {\bibinfo  {journal}
  {Advanced Quantum Technologies}\ }\textbf {\bibinfo {volume} {3}},\ \bibinfo
  {pages} {2070113} (\bibinfo {year} {2020})}\BibitemShut {NoStop}%
\bibitem [{\citenamefont {Gorman}\ \emph {et~al.}(2018)\citenamefont {Gorman},
  \citenamefont {Hemmerling}, \citenamefont {Megidish}, \citenamefont
  {Moeller}, \citenamefont {Schindler}, \citenamefont {Sarovar},\ and\
  \citenamefont {Haeffner}}]{Gorman2018}%
  \BibitemOpen
  \bibfield  {author} {\bibinfo {author} {\bibfnamefont {D.~J.}\ \bibnamefont
  {Gorman}}, \bibinfo {author} {\bibfnamefont {B.}~\bibnamefont {Hemmerling}},
  \bibinfo {author} {\bibfnamefont {E.}~\bibnamefont {Megidish}}, \bibinfo
  {author} {\bibfnamefont {S.~A.}\ \bibnamefont {Moeller}}, \bibinfo {author}
  {\bibfnamefont {P.}~\bibnamefont {Schindler}}, \bibinfo {author}
  {\bibfnamefont {M.}~\bibnamefont {Sarovar}}, \ and\ \bibinfo {author}
  {\bibfnamefont {H.}~\bibnamefont {Haeffner}},\ }\href {\doibase
  10.1103/physrevx.8.011038} {\bibfield  {journal} {\bibinfo  {journal}
  {Physical Review X}\ }\textbf {\bibinfo {volume} {8}} (\bibinfo {year}
  {2018}),\ 10.1103/physrevx.8.011038}\BibitemShut {NoStop}%
\bibitem [{\citenamefont {MacDonell}\ \emph {et~al.}(2021)\citenamefont
  {MacDonell}, \citenamefont {Dickerson}, \citenamefont {Birch}, \citenamefont
  {Kumar}, \citenamefont {Edmunds}, \citenamefont {Biercuk}, \citenamefont
  {Hempel},\ and\ \citenamefont {Kassal}}]{MacDonell2021}%
  \BibitemOpen
  \bibfield  {author} {\bibinfo {author} {\bibfnamefont {R.~J.}\ \bibnamefont
  {MacDonell}}, \bibinfo {author} {\bibfnamefont {C.~E.}\ \bibnamefont
  {Dickerson}}, \bibinfo {author} {\bibfnamefont {C.~J.~T.}\ \bibnamefont
  {Birch}}, \bibinfo {author} {\bibfnamefont {A.}~\bibnamefont {Kumar}},
  \bibinfo {author} {\bibfnamefont {C.~L.}\ \bibnamefont {Edmunds}}, \bibinfo
  {author} {\bibfnamefont {M.~J.}\ \bibnamefont {Biercuk}}, \bibinfo {author}
  {\bibfnamefont {C.}~\bibnamefont {Hempel}}, \ and\ \bibinfo {author}
  {\bibfnamefont {I.}~\bibnamefont {Kassal}},\ }\href {\doibase
  10.1039/d1sc02142g} {\bibfield  {journal} {\bibinfo  {journal} {Chemical
  Science}\ }\textbf {\bibinfo {volume} {12}},\ \bibinfo {pages} {9794}
  (\bibinfo {year} {2021})}\BibitemShut {NoStop}%
\bibitem [{\citenamefont {MacDonell}\ \emph {et~al.}(2023)\citenamefont
  {MacDonell}, \citenamefont {Navickas}, \citenamefont {Wohlers-Reichel},
  \citenamefont {Valahu}, \citenamefont {Rao}, \citenamefont {Millican},
  \citenamefont {Currington}, \citenamefont {Tan}, \citenamefont {c.~Hempel},\
  and\ \citenamefont {Kassal}}]{MacDonell2023}%
  \BibitemOpen
  \bibfield  {author} {\bibinfo {author} {\bibfnamefont {R.~J.}\ \bibnamefont
  {MacDonell}}, \bibinfo {author} {\bibfnamefont {T.}~\bibnamefont {Navickas}},
  \bibinfo {author} {\bibfnamefont {T.~F.}\ \bibnamefont {Wohlers-Reichel}},
  \bibinfo {author} {\bibfnamefont {C.~H.}\ \bibnamefont {Valahu}}, \bibinfo
  {author} {\bibfnamefont {A.~D.}\ \bibnamefont {Rao}}, \bibinfo {author}
  {\bibfnamefont {M.~J.}\ \bibnamefont {Millican}}, \bibinfo {author}
  {\bibfnamefont {M.~A.}\ \bibnamefont {Currington}}, \bibinfo {author}
  {\bibfnamefont {M.~J. B. T.~R.}\ \bibnamefont {Tan}}, \bibinfo {author}
  {\bibnamefont {c.~Hempel}}, \ and\ \bibinfo {author} {\bibfnamefont
  {I.}~\bibnamefont {Kassal}},\ }\href@noop {} {\bibfield  {journal} {\bibinfo
  {journal} {Chemical Science}\ } (\bibinfo {year} {2023})}\BibitemShut
  {NoStop}%
\bibitem [{\citenamefont {Kang}\ \emph {et~al.}(2023)\citenamefont {Kang},
  \citenamefont {Nuomin}, \citenamefont {Chowdhury}, \citenamefont {Yuly},
  \citenamefont {Sun}, \citenamefont {Whitlow}, \citenamefont {Valdiviezo},
  \citenamefont {Zhang}, \citenamefont {Zhang}, \citenamefont {Beratan},\ and\
  \citenamefont {Brown}}]{Kang2023}%
  \BibitemOpen
  \bibfield  {author} {\bibinfo {author} {\bibfnamefont {M.}~\bibnamefont
  {Kang}}, \bibinfo {author} {\bibfnamefont {H.}~\bibnamefont {Nuomin}},
  \bibinfo {author} {\bibfnamefont {S.~N.}\ \bibnamefont {Chowdhury}}, \bibinfo
  {author} {\bibfnamefont {J.~L.}\ \bibnamefont {Yuly}}, \bibinfo {author}
  {\bibfnamefont {K.}~\bibnamefont {Sun}}, \bibinfo {author} {\bibfnamefont
  {J.}~\bibnamefont {Whitlow}}, \bibinfo {author} {\bibfnamefont
  {J.}~\bibnamefont {Valdiviezo}}, \bibinfo {author} {\bibfnamefont
  {Z.}~\bibnamefont {Zhang}}, \bibinfo {author} {\bibfnamefont
  {P.}~\bibnamefont {Zhang}}, \bibinfo {author} {\bibfnamefont {D.~N.}\
  \bibnamefont {Beratan}}, \ and\ \bibinfo {author} {\bibfnamefont {K.~R.}\
  \bibnamefont {Brown}},\ }\href {\doibase 10.48550/ARXIV.2305.03156} {\enquote
  {\bibinfo {title} {Trapped-ion quantum simulations for condensed-phase
  chemical dynamics: seeking a quantum advantage},}\ } (\bibinfo {year}
  {2023})\BibitemShut {NoStop}%
\bibitem [{\citenamefont {Wang}\ \emph {et~al.}(2023)\citenamefont {Wang},
  \citenamefont {Frattini}, \citenamefont {Chapman}, \citenamefont {Puri},
  \citenamefont {Girvin}, \citenamefont {Devoret},\ and\ \citenamefont
  {Schoelkopf}}]{Wang2023}%
  \BibitemOpen
  \bibfield  {author} {\bibinfo {author} {\bibfnamefont {C.~S.}\ \bibnamefont
  {Wang}}, \bibinfo {author} {\bibfnamefont {N.~E.}\ \bibnamefont {Frattini}},
  \bibinfo {author} {\bibfnamefont {B.~J.}\ \bibnamefont {Chapman}}, \bibinfo
  {author} {\bibfnamefont {S.}~\bibnamefont {Puri}}, \bibinfo {author}
  {\bibfnamefont {S.}~\bibnamefont {Girvin}}, \bibinfo {author} {\bibfnamefont
  {M.~H.}\ \bibnamefont {Devoret}}, \ and\ \bibinfo {author} {\bibfnamefont
  {R.~J.}\ \bibnamefont {Schoelkopf}},\ }\href {\doibase
  10.1103/physrevx.13.011008} {\bibfield  {journal} {\bibinfo  {journal}
  {Physical Review X}\ }\textbf {\bibinfo {volume} {13}} (\bibinfo {year}
  {2023}),\ 10.1103/physrevx.13.011008}\BibitemShut {NoStop}%
\bibitem [{\citenamefont {Whitlow}\ \emph {et~al.}(2023)\citenamefont
  {Whitlow}, \citenamefont {Jia}, \citenamefont {Wang}, \citenamefont {Fang},
  \citenamefont {Kim},\ and\ \citenamefont {Brown}}]{Whitlow2023}%
  \BibitemOpen
  \bibfield  {author} {\bibinfo {author} {\bibfnamefont {J.}~\bibnamefont
  {Whitlow}}, \bibinfo {author} {\bibfnamefont {Z.}~\bibnamefont {Jia}},
  \bibinfo {author} {\bibfnamefont {Y.}~\bibnamefont {Wang}}, \bibinfo {author}
  {\bibfnamefont {C.}~\bibnamefont {Fang}}, \bibinfo {author} {\bibfnamefont
  {J.}~\bibnamefont {Kim}}, \ and\ \bibinfo {author} {\bibfnamefont {K.~R.}\
  \bibnamefont {Brown}},\ }\href {\doibase 10.1038/s41557-023-01303-0}
  {\bibfield  {journal} {\bibinfo  {journal} {Nature Chemistry}\ } (\bibinfo
  {year} {2023}),\ 10.1038/s41557-023-01303-0}\BibitemShut {NoStop}%
\bibitem [{\citenamefont {Valahu}\ \emph {et~al.}(2023)\citenamefont {Valahu},
  \citenamefont {Olaya-Agudelo}, \citenamefont {MacDonell}, \citenamefont
  {Navickas}, \citenamefont {Rao}, \citenamefont {Millican}, \citenamefont
  {P{\'e}rez-S{\'a}nchez}, \citenamefont {Yuen-Zhou}, \citenamefont {Biercuk},
  \citenamefont {Hempel}, \citenamefont {Tan},\ and\ \citenamefont
  {Kassal}}]{Valahu2023}%
  \BibitemOpen
  \bibfield  {author} {\bibinfo {author} {\bibfnamefont {C.~H.}\ \bibnamefont
  {Valahu}}, \bibinfo {author} {\bibfnamefont {V.~C.}\ \bibnamefont
  {Olaya-Agudelo}}, \bibinfo {author} {\bibfnamefont {R.~J.}\ \bibnamefont
  {MacDonell}}, \bibinfo {author} {\bibfnamefont {T.}~\bibnamefont {Navickas}},
  \bibinfo {author} {\bibfnamefont {A.~D.}\ \bibnamefont {Rao}}, \bibinfo
  {author} {\bibfnamefont {M.~J.}\ \bibnamefont {Millican}}, \bibinfo {author}
  {\bibfnamefont {J.~B.}\ \bibnamefont {P{\'e}rez-S{\'a}nchez}}, \bibinfo
  {author} {\bibfnamefont {J.}~\bibnamefont {Yuen-Zhou}}, \bibinfo {author}
  {\bibfnamefont {M.~J.}\ \bibnamefont {Biercuk}}, \bibinfo {author}
  {\bibfnamefont {C.}~\bibnamefont {Hempel}}, \bibinfo {author} {\bibfnamefont
  {T.~R.}\ \bibnamefont {Tan}}, \ and\ \bibinfo {author} {\bibfnamefont
  {I.}~\bibnamefont {Kassal}},\ }\href {\doibase 10.1038/s41557-023-01300-3}
  {\bibfield  {journal} {\bibinfo  {journal} {Nature Chemistry}\ } (\bibinfo
  {year} {2023}),\ 10.1038/s41557-023-01300-3}\BibitemShut {NoStop}%
\bibitem [{\citenamefont {Pham}\ \emph {et~al.}(2024)\citenamefont {Pham},
  \citenamefont {Jee}, \citenamefont {Rischka}, \citenamefont {Biercuk},\ and\
  \citenamefont {Wolf}}]{pham2024insitutunable}%
  \BibitemOpen
  \bibfield  {author} {\bibinfo {author} {\bibfnamefont {J.~H.}\ \bibnamefont
  {Pham}}, \bibinfo {author} {\bibfnamefont {J.~Y.~Z.}\ \bibnamefont {Jee}},
  \bibinfo {author} {\bibfnamefont {A.}~\bibnamefont {Rischka}}, \bibinfo
  {author} {\bibfnamefont {M.~J.}\ \bibnamefont {Biercuk}}, \ and\ \bibinfo
  {author} {\bibfnamefont {R.~N.}\ \bibnamefont {Wolf}},\ }\href@noop {}
  {\enquote {\bibinfo {title} {In-situ-tunable spin-spin interactions in a
  penning trap with in-bore optomechanics},}\ } (\bibinfo {year} {2024}),\
  \Eprint {http://arxiv.org/abs/2401.17742} {arXiv:2401.17742 [quant-ph]}
  \BibitemShut {NoStop}%
\bibitem [{\citenamefont {Biercuk}\ \emph {et~al.}(2010)\citenamefont
  {Biercuk}, \citenamefont {Uys}, \citenamefont {Britton}, \citenamefont
  {VanDevender},\ and\ \citenamefont {Bollinger}}]{biercuknaturenano2011}%
  \BibitemOpen
  \bibfield  {author} {\bibinfo {author} {\bibfnamefont {M.~J.}\ \bibnamefont
  {Biercuk}}, \bibinfo {author} {\bibfnamefont {H.}~\bibnamefont {Uys}},
  \bibinfo {author} {\bibfnamefont {J.~W.}\ \bibnamefont {Britton}}, \bibinfo
  {author} {\bibfnamefont {A.~P.}\ \bibnamefont {VanDevender}}, \ and\ \bibinfo
  {author} {\bibfnamefont {J.~J.}\ \bibnamefont {Bollinger}},\ }\href {\doibase
  10.1038/nnano.2010.165} {\bibfield  {journal} {\bibinfo  {journal} {Nature
  Nanotechnology}\ }\textbf {\bibinfo {volume} {5}},\ \bibinfo {pages} {646}
  (\bibinfo {year} {2010})}\BibitemShut {NoStop}%
\bibitem [{\citenamefont {Wolf}\ \emph {et~al.}(2019)\citenamefont {Wolf},
  \citenamefont {Shi}, \citenamefont {Heip}, \citenamefont {Gessner},
  \citenamefont {Pezz{\`{e}}}, \citenamefont {Smerzi}, \citenamefont {Schulte},
  \citenamefont {Hammerer},\ and\ \citenamefont {Schmidt}}]{Wolf2019}%
  \BibitemOpen
  \bibfield  {author} {\bibinfo {author} {\bibfnamefont {F.}~\bibnamefont
  {Wolf}}, \bibinfo {author} {\bibfnamefont {C.}~\bibnamefont {Shi}}, \bibinfo
  {author} {\bibfnamefont {J.~C.}\ \bibnamefont {Heip}}, \bibinfo {author}
  {\bibfnamefont {M.}~\bibnamefont {Gessner}}, \bibinfo {author} {\bibfnamefont
  {L.}~\bibnamefont {Pezz{\`{e}}}}, \bibinfo {author} {\bibfnamefont
  {A.}~\bibnamefont {Smerzi}}, \bibinfo {author} {\bibfnamefont
  {M.}~\bibnamefont {Schulte}}, \bibinfo {author} {\bibfnamefont
  {K.}~\bibnamefont {Hammerer}}, \ and\ \bibinfo {author} {\bibfnamefont
  {P.~O.}\ \bibnamefont {Schmidt}},\ }\href {\doibase
  10.1038/s41467-019-10576-4} {\bibfield  {journal} {\bibinfo  {journal}
  {Nature Communications}\ }\textbf {\bibinfo {volume} {10}} (\bibinfo {year}
  {2019}),\ 10.1038/s41467-019-10576-4}\BibitemShut {NoStop}%
\bibitem [{\citenamefont {McCormick}\ \emph {et~al.}(2019)\citenamefont
  {McCormick}, \citenamefont {Keller}, \citenamefont {Burd}, \citenamefont
  {Wineland}, \citenamefont {Wilson},\ and\ \citenamefont
  {Leibfried}}]{McCormick2019}%
  \BibitemOpen
  \bibfield  {author} {\bibinfo {author} {\bibfnamefont {K.~C.}\ \bibnamefont
  {McCormick}}, \bibinfo {author} {\bibfnamefont {J.}~\bibnamefont {Keller}},
  \bibinfo {author} {\bibfnamefont {S.~C.}\ \bibnamefont {Burd}}, \bibinfo
  {author} {\bibfnamefont {D.~J.}\ \bibnamefont {Wineland}}, \bibinfo {author}
  {\bibfnamefont {A.~C.}\ \bibnamefont {Wilson}}, \ and\ \bibinfo {author}
  {\bibfnamefont {D.}~\bibnamefont {Leibfried}},\ }\href {\doibase
  10.1038/s41586-019-1421-y} {\bibfield  {journal} {\bibinfo  {journal}
  {Nature}\ }\textbf {\bibinfo {volume} {572}},\ \bibinfo {pages} {86}
  (\bibinfo {year} {2019})}\BibitemShut {NoStop}%
\bibitem [{\citenamefont {Gheorghiu}\ \emph {et~al.}(2018)\citenamefont
  {Gheorghiu}, \citenamefont {Kapourniotis},\ and\ \citenamefont
  {Kashefi}}]{Gheorghiu2018}%
  \BibitemOpen
  \bibfield  {author} {\bibinfo {author} {\bibfnamefont {A.}~\bibnamefont
  {Gheorghiu}}, \bibinfo {author} {\bibfnamefont {T.}~\bibnamefont
  {Kapourniotis}}, \ and\ \bibinfo {author} {\bibfnamefont {E.}~\bibnamefont
  {Kashefi}},\ }\href {\doibase 10.1007/s00224-018-9872-3} {\bibfield
  {journal} {\bibinfo  {journal} {Theory of Computing Systems}\ }\textbf
  {\bibinfo {volume} {63}},\ \bibinfo {pages} {715–808} (\bibinfo {year}
  {2018})}\BibitemShut {NoStop}%
\bibitem [{\citenamefont {\v{S}upi\'{c}}\ and\ \citenamefont
  {Bowles}(2020)}]{supic2020}%
  \BibitemOpen
  \bibfield  {author} {\bibinfo {author} {\bibfnamefont {I.}~\bibnamefont
  {\v{S}upi\'{c}}}\ and\ \bibinfo {author} {\bibfnamefont {J.}~\bibnamefont
  {Bowles}},\ }\href {\doibase 10.22331/q-2020-09-30-337} {\bibfield  {journal}
  {\bibinfo  {journal} {Quantum}\ }\textbf {\bibinfo {volume} {4}},\ \bibinfo
  {pages} {337} (\bibinfo {year} {2020})}\BibitemShut {NoStop}%
\bibitem [{\citenamefont {Eisert}\ \emph {et~al.}(2020)\citenamefont {Eisert},
  \citenamefont {Hangleiter}, \citenamefont {Walk}, \citenamefont {Roth},
  \citenamefont {Markham}, \citenamefont {Parekh}, \citenamefont {Chabaud},\
  and\ \citenamefont {Kashefi}}]{Eisert2020}%
  \BibitemOpen
  \bibfield  {author} {\bibinfo {author} {\bibfnamefont {J.}~\bibnamefont
  {Eisert}}, \bibinfo {author} {\bibfnamefont {D.}~\bibnamefont {Hangleiter}},
  \bibinfo {author} {\bibfnamefont {N.}~\bibnamefont {Walk}}, \bibinfo {author}
  {\bibfnamefont {I.}~\bibnamefont {Roth}}, \bibinfo {author} {\bibfnamefont
  {D.}~\bibnamefont {Markham}}, \bibinfo {author} {\bibfnamefont
  {R.}~\bibnamefont {Parekh}}, \bibinfo {author} {\bibfnamefont
  {U.}~\bibnamefont {Chabaud}}, \ and\ \bibinfo {author} {\bibfnamefont
  {E.}~\bibnamefont {Kashefi}},\ }\href {\doibase 10.1038/s42254-020-0186-4}
  {\bibfield  {journal} {\bibinfo  {journal} {Nature Reviews Physics}\ }\textbf
  {\bibinfo {volume} {2}},\ \bibinfo {pages} {382–390} (\bibinfo {year}
  {2020})}\BibitemShut {NoStop}%
\bibitem [{\citenamefont {Ball}\ \emph {et~al.}(2021)\citenamefont {Ball},
  \citenamefont {Biercuk}, \citenamefont {Carvalho}, \citenamefont {Chen},
  \citenamefont {Hush}, \citenamefont {Castro}, \citenamefont {Li},
  \citenamefont {Liebermann}, \citenamefont {Slatyer}, \citenamefont {Edmunds},
  \citenamefont {Frey}, \citenamefont {Hempel},\ and\ \citenamefont
  {Milne}}]{Ball_2021}%
  \BibitemOpen
  \bibfield  {author} {\bibinfo {author} {\bibfnamefont {H.}~\bibnamefont
  {Ball}}, \bibinfo {author} {\bibfnamefont {M.~J.}\ \bibnamefont {Biercuk}},
  \bibinfo {author} {\bibfnamefont {A.~R.~R.}\ \bibnamefont {Carvalho}},
  \bibinfo {author} {\bibfnamefont {J.}~\bibnamefont {Chen}}, \bibinfo {author}
  {\bibfnamefont {M.}~\bibnamefont {Hush}}, \bibinfo {author} {\bibfnamefont
  {L.~A.~D.}\ \bibnamefont {Castro}}, \bibinfo {author} {\bibfnamefont
  {L.}~\bibnamefont {Li}}, \bibinfo {author} {\bibfnamefont {P.~J.}\
  \bibnamefont {Liebermann}}, \bibinfo {author} {\bibfnamefont {H.~J.}\
  \bibnamefont {Slatyer}}, \bibinfo {author} {\bibfnamefont {C.}~\bibnamefont
  {Edmunds}}, \bibinfo {author} {\bibfnamefont {V.}~\bibnamefont {Frey}},
  \bibinfo {author} {\bibfnamefont {C.}~\bibnamefont {Hempel}}, \ and\ \bibinfo
  {author} {\bibfnamefont {A.}~\bibnamefont {Milne}},\ }\href {\doibase
  10.1088/2058-9565/abdca6} {\bibfield  {journal} {\bibinfo  {journal} {Quantum
  Science and Technology}\ }\textbf {\bibinfo {volume} {6}},\ \bibinfo {pages}
  {044011} (\bibinfo {year} {2021})}\BibitemShut {NoStop}%
\bibitem [{\citenamefont {Emerson}\ \emph {et~al.}(2005)\citenamefont
  {Emerson}, \citenamefont {Alicki},\ and\ \citenamefont
  {{\.{Z}}yczkowski}}]{Emerson2005}%
  \BibitemOpen
  \bibfield  {author} {\bibinfo {author} {\bibfnamefont {J.}~\bibnamefont
  {Emerson}}, \bibinfo {author} {\bibfnamefont {R.}~\bibnamefont {Alicki}}, \
  and\ \bibinfo {author} {\bibfnamefont {K.}~\bibnamefont {{\.{Z}}yczkowski}},\
  }\href {\doibase 10.1088/1464-4266/7/10/021} {\bibfield  {journal} {\bibinfo
  {journal} {Journal of Optics B: Quantum and Semiclassical Optics}\ }\textbf
  {\bibinfo {volume} {7}},\ \bibinfo {pages} {S347} (\bibinfo {year}
  {2005})}\BibitemShut {NoStop}%
\bibitem [{\citenamefont {Magesan}\ \emph {et~al.}(2011)\citenamefont
  {Magesan}, \citenamefont {Gambetta},\ and\ \citenamefont
  {Emerson}}]{Magesan2011}%
  \BibitemOpen
  \bibfield  {author} {\bibinfo {author} {\bibfnamefont {E.}~\bibnamefont
  {Magesan}}, \bibinfo {author} {\bibfnamefont {J.~M.}\ \bibnamefont
  {Gambetta}}, \ and\ \bibinfo {author} {\bibfnamefont {J.}~\bibnamefont
  {Emerson}},\ }\href {\doibase 10.1103/physrevlett.106.180504} {\bibfield
  {journal} {\bibinfo  {journal} {Physical Review Letters}\ }\textbf {\bibinfo
  {volume} {106}} (\bibinfo {year} {2011}),\
  10.1103/physrevlett.106.180504}\BibitemShut {NoStop}%
\bibitem [{\citenamefont {Magesan}\ \emph {et~al.}(2012)\citenamefont
  {Magesan}, \citenamefont {Gambetta}, \citenamefont {Johnson}, \citenamefont
  {Ryan}, \citenamefont {Chow}, \citenamefont {Merkel}, \citenamefont
  {da~Silva}, \citenamefont {Keefe}, \citenamefont {Rothwell}, \citenamefont
  {Ohki}, \citenamefont {Ketchen},\ and\ \citenamefont
  {Steffen}}]{Magesan2012}%
  \BibitemOpen
  \bibfield  {author} {\bibinfo {author} {\bibfnamefont {E.}~\bibnamefont
  {Magesan}}, \bibinfo {author} {\bibfnamefont {J.~M.}\ \bibnamefont
  {Gambetta}}, \bibinfo {author} {\bibfnamefont {B.~R.}\ \bibnamefont
  {Johnson}}, \bibinfo {author} {\bibfnamefont {C.~A.}\ \bibnamefont {Ryan}},
  \bibinfo {author} {\bibfnamefont {J.~M.}\ \bibnamefont {Chow}}, \bibinfo
  {author} {\bibfnamefont {S.~T.}\ \bibnamefont {Merkel}}, \bibinfo {author}
  {\bibfnamefont {M.~P.}\ \bibnamefont {da~Silva}}, \bibinfo {author}
  {\bibfnamefont {G.~A.}\ \bibnamefont {Keefe}}, \bibinfo {author}
  {\bibfnamefont {M.~B.}\ \bibnamefont {Rothwell}}, \bibinfo {author}
  {\bibfnamefont {T.~A.}\ \bibnamefont {Ohki}}, \bibinfo {author}
  {\bibfnamefont {M.~B.}\ \bibnamefont {Ketchen}}, \ and\ \bibinfo {author}
  {\bibfnamefont {M.}~\bibnamefont {Steffen}},\ }\href {\doibase
  10.1103/physrevlett.109.080505} {\bibfield  {journal} {\bibinfo  {journal}
  {Physical Review Letters}\ }\textbf {\bibinfo {volume} {109}} (\bibinfo
  {year} {2012}),\ 10.1103/physrevlett.109.080505}\BibitemShut {NoStop}%
\bibitem [{\citenamefont {Epstein}\ \emph {et~al.}(2014)\citenamefont
  {Epstein}, \citenamefont {Cross}, \citenamefont {Magesan},\ and\
  \citenamefont {Gambetta}}]{Epstein2014}%
  \BibitemOpen
  \bibfield  {author} {\bibinfo {author} {\bibfnamefont {J.~M.}\ \bibnamefont
  {Epstein}}, \bibinfo {author} {\bibfnamefont {A.~W.}\ \bibnamefont {Cross}},
  \bibinfo {author} {\bibfnamefont {E.}~\bibnamefont {Magesan}}, \ and\
  \bibinfo {author} {\bibfnamefont {J.~M.}\ \bibnamefont {Gambetta}},\ }\href
  {\doibase 10.1103/physreva.89.062321} {\bibfield  {journal} {\bibinfo
  {journal} {Physical Review A}\ }\textbf {\bibinfo {volume} {89}} (\bibinfo
  {year} {2014}),\ 10.1103/physreva.89.062321}\BibitemShut {NoStop}%
\bibitem [{\citenamefont {Ball}\ \emph {et~al.}(2016)\citenamefont {Ball},
  \citenamefont {Stace}, \citenamefont {Flammia},\ and\ \citenamefont
  {Biercuk}}]{Ball2016}%
  \BibitemOpen
  \bibfield  {author} {\bibinfo {author} {\bibfnamefont {H.}~\bibnamefont
  {Ball}}, \bibinfo {author} {\bibfnamefont {T.~M.}\ \bibnamefont {Stace}},
  \bibinfo {author} {\bibfnamefont {S.~T.}\ \bibnamefont {Flammia}}, \ and\
  \bibinfo {author} {\bibfnamefont {M.~J.}\ \bibnamefont {Biercuk}},\ }\href
  {\doibase 10.1103/physreva.93.022303} {\bibfield  {journal} {\bibinfo
  {journal} {Physical Review A}\ }\textbf {\bibinfo {volume} {93}} (\bibinfo
  {year} {2016}),\ 10.1103/physreva.93.022303}\BibitemShut {NoStop}%
\bibitem [{\citenamefont {Kueng}\ \emph {et~al.}(2016)\citenamefont {Kueng},
  \citenamefont {Long}, \citenamefont {Doherty},\ and\ \citenamefont
  {Flammia}}]{Keung2016}%
  \BibitemOpen
  \bibfield  {author} {\bibinfo {author} {\bibfnamefont {R.}~\bibnamefont
  {Kueng}}, \bibinfo {author} {\bibfnamefont {D.~M.}\ \bibnamefont {Long}},
  \bibinfo {author} {\bibfnamefont {A.~C.}\ \bibnamefont {Doherty}}, \ and\
  \bibinfo {author} {\bibfnamefont {S.~T.}\ \bibnamefont {Flammia}},\ }\href
  {\doibase 10.1103/PhysRevLett.117.170502} {\bibfield  {journal} {\bibinfo
  {journal} {Phys. Rev. Lett.}\ }\textbf {\bibinfo {volume} {117}},\ \bibinfo
  {pages} {170502} (\bibinfo {year} {2016})}\BibitemShut {NoStop}%
\bibitem [{\citenamefont {Mavadia}\ \emph {et~al.}(2018)\citenamefont
  {Mavadia}, \citenamefont {Edmunds}, \citenamefont {Hempel}, \citenamefont
  {Ball}, \citenamefont {Roy}, \citenamefont {Stace},\ and\ \citenamefont
  {Biercuk}}]{Mavadia2018}%
  \BibitemOpen
  \bibfield  {author} {\bibinfo {author} {\bibfnamefont {S.}~\bibnamefont
  {Mavadia}}, \bibinfo {author} {\bibfnamefont {C.~L.}\ \bibnamefont
  {Edmunds}}, \bibinfo {author} {\bibfnamefont {C.}~\bibnamefont {Hempel}},
  \bibinfo {author} {\bibfnamefont {H.}~\bibnamefont {Ball}}, \bibinfo {author}
  {\bibfnamefont {F.}~\bibnamefont {Roy}}, \bibinfo {author} {\bibfnamefont
  {T.~M.}\ \bibnamefont {Stace}}, \ and\ \bibinfo {author} {\bibfnamefont
  {M.~J.}\ \bibnamefont {Biercuk}},\ }\href {\doibase
  10.1038/s41534-017-0052-0} {\bibfield  {journal} {\bibinfo  {journal} {npj
  Quantum Information}\ }\textbf {\bibinfo {volume} {4}},\ \bibinfo {pages} {7}
  (\bibinfo {year} {2018})}\BibitemShut {NoStop}%
\bibitem [{\citenamefont {Edmunds}\ \emph {et~al.}(2020)\citenamefont
  {Edmunds}, \citenamefont {Hempel}, \citenamefont {Harris}, \citenamefont
  {Frey}, \citenamefont {Stace},\ and\ \citenamefont {Biercuk}}]{Edmunds2020}%
  \BibitemOpen
  \bibfield  {author} {\bibinfo {author} {\bibfnamefont {C.~L.}\ \bibnamefont
  {Edmunds}}, \bibinfo {author} {\bibfnamefont {C.}~\bibnamefont {Hempel}},
  \bibinfo {author} {\bibfnamefont {R.~J.}\ \bibnamefont {Harris}}, \bibinfo
  {author} {\bibfnamefont {V.}~\bibnamefont {Frey}}, \bibinfo {author}
  {\bibfnamefont {T.~M.}\ \bibnamefont {Stace}}, \ and\ \bibinfo {author}
  {\bibfnamefont {M.~J.}\ \bibnamefont {Biercuk}},\ }\href {\doibase
  10.1103/PhysRevResearch.2.013156} {\bibfield  {journal} {\bibinfo  {journal}
  {Phys. Rev. Res.}\ }\textbf {\bibinfo {volume} {2}},\ \bibinfo {pages}
  {013156} (\bibinfo {year} {2020})}\BibitemShut {NoStop}%
\bibitem [{\citenamefont {Elben}\ \emph {et~al.}(2020)\citenamefont {Elben},
  \citenamefont {Vermersch}, \citenamefont {van Bijnen}, \citenamefont
  {Kokail}, \citenamefont {Brydges}, \citenamefont {Maier}, \citenamefont
  {Joshi}, \citenamefont {Blatt}, \citenamefont {Roos},\ and\ \citenamefont
  {Zoller}}]{Elben2020}%
  \BibitemOpen
  \bibfield  {author} {\bibinfo {author} {\bibfnamefont {A.}~\bibnamefont
  {Elben}}, \bibinfo {author} {\bibfnamefont {B.}~\bibnamefont {Vermersch}},
  \bibinfo {author} {\bibfnamefont {R.}~\bibnamefont {van Bijnen}}, \bibinfo
  {author} {\bibfnamefont {C.}~\bibnamefont {Kokail}}, \bibinfo {author}
  {\bibfnamefont {T.}~\bibnamefont {Brydges}}, \bibinfo {author} {\bibfnamefont
  {C.}~\bibnamefont {Maier}}, \bibinfo {author} {\bibfnamefont {M.~K.}\
  \bibnamefont {Joshi}}, \bibinfo {author} {\bibfnamefont {R.}~\bibnamefont
  {Blatt}}, \bibinfo {author} {\bibfnamefont {C.~F.}\ \bibnamefont {Roos}}, \
  and\ \bibinfo {author} {\bibfnamefont {P.}~\bibnamefont {Zoller}},\ }\href
  {\doibase 10.1103/physrevlett.124.010504} {\bibfield  {journal} {\bibinfo
  {journal} {Physical Review Letters}\ }\textbf {\bibinfo {volume} {124}}
  (\bibinfo {year} {2020}),\ 10.1103/physrevlett.124.010504}\BibitemShut
  {NoStop}%
\bibitem [{\citenamefont {Shaffer}\ \emph {et~al.}(2023)\citenamefont
  {Shaffer}, \citenamefont {Ren}, \citenamefont {Dyrenkova}, \citenamefont
  {Yale}, \citenamefont {Lobser}, \citenamefont {Burch}, \citenamefont {Chow},
  \citenamefont {Revelle}, \citenamefont {Clark},\ and\ \citenamefont
  {H\"{a}ffner}}]{Shaffer2023}%
  \BibitemOpen
  \bibfield  {author} {\bibinfo {author} {\bibfnamefont {R.}~\bibnamefont
  {Shaffer}}, \bibinfo {author} {\bibfnamefont {H.}~\bibnamefont {Ren}},
  \bibinfo {author} {\bibfnamefont {E.}~\bibnamefont {Dyrenkova}}, \bibinfo
  {author} {\bibfnamefont {C.~G.}\ \bibnamefont {Yale}}, \bibinfo {author}
  {\bibfnamefont {D.~S.}\ \bibnamefont {Lobser}}, \bibinfo {author}
  {\bibfnamefont {A.~D.}\ \bibnamefont {Burch}}, \bibinfo {author}
  {\bibfnamefont {M.~N.~H.}\ \bibnamefont {Chow}}, \bibinfo {author}
  {\bibfnamefont {M.~C.}\ \bibnamefont {Revelle}}, \bibinfo {author}
  {\bibfnamefont {S.~M.}\ \bibnamefont {Clark}}, \ and\ \bibinfo {author}
  {\bibfnamefont {H.}~\bibnamefont {H\"{a}ffner}},\ }\href {\doibase
  10.22331/q-2023-05-04-997} {\bibfield  {journal} {\bibinfo  {journal}
  {Quantum}\ }\textbf {\bibinfo {volume} {7}},\ \bibinfo {pages} {997}
  (\bibinfo {year} {2023})}\BibitemShut {NoStop}%
\bibitem [{\citenamefont {Derbyshire}\ \emph {et~al.}(2020)\citenamefont
  {Derbyshire}, \citenamefont {Malo}, \citenamefont {Daley}, \citenamefont
  {Kashefi},\ and\ \citenamefont {Wallden}}]{Derbyshire2020}%
  \BibitemOpen
  \bibfield  {author} {\bibinfo {author} {\bibfnamefont {E.}~\bibnamefont
  {Derbyshire}}, \bibinfo {author} {\bibfnamefont {J.~Y.}\ \bibnamefont
  {Malo}}, \bibinfo {author} {\bibfnamefont {A.~J.}\ \bibnamefont {Daley}},
  \bibinfo {author} {\bibfnamefont {E.}~\bibnamefont {Kashefi}}, \ and\
  \bibinfo {author} {\bibfnamefont {P.}~\bibnamefont {Wallden}},\ }\href
  {\doibase 10.1088/2058-9565/ab7eec} {\bibfield  {journal} {\bibinfo
  {journal} {Quantum Science and Technology}\ }\textbf {\bibinfo {volume}
  {5}},\ \bibinfo {pages} {034001} (\bibinfo {year} {2020})}\BibitemShut
  {NoStop}%
\bibitem [{\citenamefont {Shaffer}\ \emph {et~al.}(2021)\citenamefont
  {Shaffer}, \citenamefont {Megidish}, \citenamefont {Broz}, \citenamefont
  {Chen},\ and\ \citenamefont {Häffner}}]{Shaffer2021}%
  \BibitemOpen
  \bibfield  {author} {\bibinfo {author} {\bibfnamefont {R.}~\bibnamefont
  {Shaffer}}, \bibinfo {author} {\bibfnamefont {E.}~\bibnamefont {Megidish}},
  \bibinfo {author} {\bibfnamefont {J.}~\bibnamefont {Broz}}, \bibinfo {author}
  {\bibfnamefont {W.-T.}\ \bibnamefont {Chen}}, \ and\ \bibinfo {author}
  {\bibfnamefont {H.}~\bibnamefont {Häffner}},\ }\href {\doibase
  10.1038/s41534-021-00380-8} {\bibfield  {journal} {\bibinfo  {journal} {npj
  Quantum Information}\ }\textbf {\bibinfo {volume} {7}} (\bibinfo {year}
  {2021}),\ 10.1038/s41534-021-00380-8}\BibitemShut {NoStop}%
\bibitem [{\citenamefont {Wu}\ and\ \citenamefont {Sanders}(2019)}]{Wu2019}%
  \BibitemOpen
  \bibfield  {author} {\bibinfo {author} {\bibfnamefont {Y.-D.}\ \bibnamefont
  {Wu}}\ and\ \bibinfo {author} {\bibfnamefont {B.~C.}\ \bibnamefont
  {Sanders}},\ }\href {\doibase 10.1088/1367-2630/ab2d3a} {\bibfield  {journal}
  {\bibinfo  {journal} {New Journal of Physics}\ }\textbf {\bibinfo {volume}
  {21}},\ \bibinfo {pages} {073026} (\bibinfo {year} {2019})}\BibitemShut
  {NoStop}%
\bibitem [{\citenamefont {Roos}\ \emph {et~al.}(2008)\citenamefont {Roos},
  \citenamefont {Monz}, \citenamefont {Kim}, \citenamefont {Riebe},
  \citenamefont {H\"{a}ffner}, \citenamefont {James},\ and\ \citenamefont
  {Blatt}}]{Roos2008}%
  \BibitemOpen
  \bibfield  {author} {\bibinfo {author} {\bibfnamefont {C.~F.}\ \bibnamefont
  {Roos}}, \bibinfo {author} {\bibfnamefont {T.}~\bibnamefont {Monz}}, \bibinfo
  {author} {\bibfnamefont {K.}~\bibnamefont {Kim}}, \bibinfo {author}
  {\bibfnamefont {M.}~\bibnamefont {Riebe}}, \bibinfo {author} {\bibfnamefont
  {H.}~\bibnamefont {H\"{a}ffner}}, \bibinfo {author} {\bibfnamefont
  {D.~F.~V.}\ \bibnamefont {James}}, \ and\ \bibinfo {author} {\bibfnamefont
  {R.}~\bibnamefont {Blatt}},\ }\href {\doibase 10.1103/physreva.77.040302}
  {\bibfield  {journal} {\bibinfo  {journal} {Physical Review A}\ }\textbf
  {\bibinfo {volume} {77}} (\bibinfo {year} {2008}),\
  10.1103/physreva.77.040302}\BibitemShut {NoStop}%
\bibitem [{\citenamefont {Brownnutt}\ \emph {et~al.}(2015)\citenamefont
  {Brownnutt}, \citenamefont {Kumph}, \citenamefont {Rabl},\ and\ \citenamefont
  {Blatt}}]{Brownnutt2015}%
  \BibitemOpen
  \bibfield  {author} {\bibinfo {author} {\bibfnamefont {M.}~\bibnamefont
  {Brownnutt}}, \bibinfo {author} {\bibfnamefont {M.}~\bibnamefont {Kumph}},
  \bibinfo {author} {\bibfnamefont {P.}~\bibnamefont {Rabl}}, \ and\ \bibinfo
  {author} {\bibfnamefont {R.}~\bibnamefont {Blatt}},\ }\href {\doibase
  10.1103/revmodphys.87.1419} {\bibfield  {journal} {\bibinfo  {journal}
  {Reviews of Modern Physics}\ }\textbf {\bibinfo {volume} {87}},\ \bibinfo
  {pages} {1419–1482} (\bibinfo {year} {2015})}\BibitemShut {NoStop}%
\bibitem [{\citenamefont {Milne}\ \emph {et~al.}(2021)\citenamefont {Milne},
  \citenamefont {Hempel}, \citenamefont {Li}, \citenamefont {Edmunds},
  \citenamefont {Slatyer}, \citenamefont {Ball}, \citenamefont {Hush},\ and\
  \citenamefont {Biercuk}}]{Milne2021}%
  \BibitemOpen
  \bibfield  {author} {\bibinfo {author} {\bibfnamefont {A.~R.}\ \bibnamefont
  {Milne}}, \bibinfo {author} {\bibfnamefont {C.}~\bibnamefont {Hempel}},
  \bibinfo {author} {\bibfnamefont {L.}~\bibnamefont {Li}}, \bibinfo {author}
  {\bibfnamefont {C.~L.}\ \bibnamefont {Edmunds}}, \bibinfo {author}
  {\bibfnamefont {H.~J.}\ \bibnamefont {Slatyer}}, \bibinfo {author}
  {\bibfnamefont {H.}~\bibnamefont {Ball}}, \bibinfo {author} {\bibfnamefont
  {M.~R.}\ \bibnamefont {Hush}}, \ and\ \bibinfo {author} {\bibfnamefont
  {M.~J.}\ \bibnamefont {Biercuk}},\ }\href {\doibase
  10.1103/PhysRevLett.126.250506} {\bibfield  {journal} {\bibinfo  {journal}
  {Phys. Rev. Lett.}\ }\textbf {\bibinfo {volume} {126}},\ \bibinfo {pages}
  {250506} (\bibinfo {year} {2021})}\BibitemShut {NoStop}%
\bibitem [{\citenamefont {Fogarty}\ \emph {et~al.}(2015)\citenamefont
  {Fogarty}, \citenamefont {Veldhorst}, \citenamefont {Harper}, \citenamefont
  {Yang}, \citenamefont {Bartlett}, \citenamefont {Flammia},\ and\
  \citenamefont {Dzurak}}]{Fogarty2015}%
  \BibitemOpen
  \bibfield  {author} {\bibinfo {author} {\bibfnamefont {M.~A.}\ \bibnamefont
  {Fogarty}}, \bibinfo {author} {\bibfnamefont {M.}~\bibnamefont {Veldhorst}},
  \bibinfo {author} {\bibfnamefont {R.}~\bibnamefont {Harper}}, \bibinfo
  {author} {\bibfnamefont {C.~H.}\ \bibnamefont {Yang}}, \bibinfo {author}
  {\bibfnamefont {S.~D.}\ \bibnamefont {Bartlett}}, \bibinfo {author}
  {\bibfnamefont {S.~T.}\ \bibnamefont {Flammia}}, \ and\ \bibinfo {author}
  {\bibfnamefont {A.~S.}\ \bibnamefont {Dzurak}},\ }\href {\doibase
  10.1103/physreva.92.022326} {\bibfield  {journal} {\bibinfo  {journal}
  {Physical Review A}\ }\textbf {\bibinfo {volume} {92}} (\bibinfo {year}
  {2015}),\ 10.1103/physreva.92.022326}\BibitemShut {NoStop}%
\bibitem [{\citenamefont {Figueroa-Romero}\ \emph {et~al.}(2021)\citenamefont
  {Figueroa-Romero}, \citenamefont {Modi}, \citenamefont {Harris},
  \citenamefont {Stace},\ and\ \citenamefont {Hsieh}}]{FigueroaRomero2021}%
  \BibitemOpen
  \bibfield  {author} {\bibinfo {author} {\bibfnamefont {P.}~\bibnamefont
  {Figueroa-Romero}}, \bibinfo {author} {\bibfnamefont {K.}~\bibnamefont
  {Modi}}, \bibinfo {author} {\bibfnamefont {R.~J.}\ \bibnamefont {Harris}},
  \bibinfo {author} {\bibfnamefont {T.~M.}\ \bibnamefont {Stace}}, \ and\
  \bibinfo {author} {\bibfnamefont {M.-H.}\ \bibnamefont {Hsieh}},\ }\href
  {\doibase 10.1103/prxquantum.2.040351} {\bibfield  {journal} {\bibinfo
  {journal} {PRX Quantum}\ }\textbf {\bibinfo {volume} {2}} (\bibinfo {year}
  {2021}),\ 10.1103/prxquantum.2.040351}\BibitemShut {NoStop}%
\bibitem [{\citenamefont {James}(1998)}]{James1998b}%
  \BibitemOpen
  \bibfield  {author} {\bibinfo {author} {\bibfnamefont {D.~F.~V.}\
  \bibnamefont {James}},\ }\href {\doibase 10.1103/physrevlett.81.317}
  {\bibfield  {journal} {\bibinfo  {journal} {Physical Review Letters}\
  }\textbf {\bibinfo {volume} {81}},\ \bibinfo {pages} {317–320} (\bibinfo
  {year} {1998})}\BibitemShut {NoStop}%
\bibitem [{\citenamefont {Rasmusson}\ \emph {et~al.}(2024)\citenamefont
  {Rasmusson}, \citenamefont {Jung}, \citenamefont {Schroer}, \citenamefont
  {Kyprianidis},\ and\ \citenamefont {Richerme}}]{rasmusson2024}%
  \BibitemOpen
  \bibfield  {author} {\bibinfo {author} {\bibfnamefont {A.~J.}\ \bibnamefont
  {Rasmusson}}, \bibinfo {author} {\bibfnamefont {I.}~\bibnamefont {Jung}},
  \bibinfo {author} {\bibfnamefont {F.}~\bibnamefont {Schroer}}, \bibinfo
  {author} {\bibfnamefont {A.}~\bibnamefont {Kyprianidis}}, \ and\ \bibinfo
  {author} {\bibfnamefont {P.}~\bibnamefont {Richerme}},\ }\href {\doibase
  10.48550/ARXIV.2404.09327} {\enquote {\bibinfo {title} {Measurement-induced
  heating of trapped ions},}\ } (\bibinfo {year} {2024})\BibitemShut {NoStop}%
\bibitem [{\citenamefont {Wineland}\ \emph {et~al.}(1998)\citenamefont
  {Wineland}, \citenamefont {Monroe}, \citenamefont {Itano}, \citenamefont
  {Leibfried}, \citenamefont {King},\ and\ \citenamefont
  {Meekhof}}]{Wineland1998}%
  \BibitemOpen
  \bibfield  {author} {\bibinfo {author} {\bibfnamefont {D.}~\bibnamefont
  {Wineland}}, \bibinfo {author} {\bibfnamefont {C.}~\bibnamefont {Monroe}},
  \bibinfo {author} {\bibfnamefont {W.}~\bibnamefont {Itano}}, \bibinfo
  {author} {\bibfnamefont {D.}~\bibnamefont {Leibfried}}, \bibinfo {author}
  {\bibfnamefont {B.}~\bibnamefont {King}}, \ and\ \bibinfo {author}
  {\bibfnamefont {D.}~\bibnamefont {Meekhof}},\ }\href {\doibase
  10.6028/jres.103.019} {\bibfield  {journal} {\bibinfo  {journal} {Journal of
  Research of the National Institute of Standards and Technology}\ }\textbf
  {\bibinfo {volume} {103}},\ \bibinfo {pages} {259} (\bibinfo {year}
  {1998})}\BibitemShut {NoStop}%
\bibitem [{\citenamefont {Monroe}\ \emph {et~al.}(1995)\citenamefont {Monroe},
  \citenamefont {Meekhof}, \citenamefont {King}, \citenamefont {Jefferts},
  \citenamefont {Itano}, \citenamefont {Wineland},\ and\ \citenamefont
  {Gould}}]{Monroe1995}%
  \BibitemOpen
  \bibfield  {author} {\bibinfo {author} {\bibfnamefont {C.}~\bibnamefont
  {Monroe}}, \bibinfo {author} {\bibfnamefont {D.~M.}\ \bibnamefont {Meekhof}},
  \bibinfo {author} {\bibfnamefont {B.~E.}\ \bibnamefont {King}}, \bibinfo
  {author} {\bibfnamefont {S.~R.}\ \bibnamefont {Jefferts}}, \bibinfo {author}
  {\bibfnamefont {W.~M.}\ \bibnamefont {Itano}}, \bibinfo {author}
  {\bibfnamefont {D.~J.}\ \bibnamefont {Wineland}}, \ and\ \bibinfo {author}
  {\bibfnamefont {P.}~\bibnamefont {Gould}},\ }\href {\doibase
  10.1103/physrevlett.75.4011} {\bibfield  {journal} {\bibinfo  {journal}
  {Physical Review Letters}\ }\textbf {\bibinfo {volume} {75}},\ \bibinfo
  {pages} {4011} (\bibinfo {year} {1995})}\BibitemShut {NoStop}%
\bibitem [{\citenamefont {Turchette}\ \emph {et~al.}(2000)\citenamefont
  {Turchette}, \citenamefont {Kielpinski}, \citenamefont {King}, \citenamefont
  {Leibfried}, \citenamefont {Meekhof}, \citenamefont {Myatt}, \citenamefont
  {Rowe}, \citenamefont {Sackett}, \citenamefont {Wood}, \citenamefont {Itano},
  \citenamefont {Monroe},\ and\ \citenamefont {Wineland}}]{Turchette2000}%
  \BibitemOpen
  \bibfield  {author} {\bibinfo {author} {\bibfnamefont {Q.~A.}\ \bibnamefont
  {Turchette}}, \bibinfo {author} {\bibnamefont {Kielpinski}}, \bibinfo
  {author} {\bibfnamefont {B.~E.}\ \bibnamefont {King}}, \bibinfo {author}
  {\bibfnamefont {D.}~\bibnamefont {Leibfried}}, \bibinfo {author}
  {\bibfnamefont {D.~M.}\ \bibnamefont {Meekhof}}, \bibinfo {author}
  {\bibfnamefont {C.~J.}\ \bibnamefont {Myatt}}, \bibinfo {author}
  {\bibfnamefont {M.~A.}\ \bibnamefont {Rowe}}, \bibinfo {author}
  {\bibfnamefont {C.~A.}\ \bibnamefont {Sackett}}, \bibinfo {author}
  {\bibfnamefont {C.~S.}\ \bibnamefont {Wood}}, \bibinfo {author}
  {\bibfnamefont {W.~M.}\ \bibnamefont {Itano}}, \bibinfo {author}
  {\bibfnamefont {C.}~\bibnamefont {Monroe}}, \ and\ \bibinfo {author}
  {\bibfnamefont {D.~J.}\ \bibnamefont {Wineland}},\ }\href {\doibase
  10.1103/physreva.61.063418} {\bibfield  {journal} {\bibinfo  {journal}
  {Physical Review A}\ }\textbf {\bibinfo {volume} {61}} (\bibinfo {year}
  {2000}),\ 10.1103/physreva.61.063418}\BibitemShut {NoStop}%
\bibitem [{\citenamefont {Efron}\ and\ \citenamefont
  {Tibshirani}(1994)}]{Efron1994}%
  \BibitemOpen
  \bibfield  {author} {\bibinfo {author} {\bibfnamefont {B.}~\bibnamefont
  {Efron}}\ and\ \bibinfo {author} {\bibfnamefont {R.}~\bibnamefont
  {Tibshirani}},\ }\href {\doibase 10.1201/9780429246593} {\emph {\bibinfo
  {title} {An Introduction to the Bootstrap}}}\ (\bibinfo  {publisher} {Chapman
  and Hall/CRC},\ \bibinfo {year} {1994})\BibitemShut {NoStop}%
\bibitem [{\citenamefont {Savard}\ \emph {et~al.}(1997)\citenamefont {Savard},
  \citenamefont {O’Hara},\ and\ \citenamefont {Thomas}}]{Savard1997}%
  \BibitemOpen
  \bibfield  {author} {\bibinfo {author} {\bibfnamefont {T.~A.}\ \bibnamefont
  {Savard}}, \bibinfo {author} {\bibfnamefont {K.~M.}\ \bibnamefont
  {O’Hara}}, \ and\ \bibinfo {author} {\bibfnamefont {J.~E.}\ \bibnamefont
  {Thomas}},\ }\href {\doibase 10.1103/physreva.56.r1095} {\bibfield  {journal}
  {\bibinfo  {journal} {Physical Review A}\ }\textbf {\bibinfo {volume} {56}},\
  \bibinfo {pages} {R1095–R1098} (\bibinfo {year} {1997})}\BibitemShut
  {NoStop}%
\bibitem [{\citenamefont {Lee}\ \emph {et~al.}(2016)\citenamefont {Lee},
  \citenamefont {Smith}, \citenamefont {Richerme}, \citenamefont {Neyenhuis},
  \citenamefont {Hess}, \citenamefont {Zhang},\ and\ \citenamefont
  {Monroe}}]{Lee2016}%
  \BibitemOpen
  \bibfield  {author} {\bibinfo {author} {\bibfnamefont {A.~C.}\ \bibnamefont
  {Lee}}, \bibinfo {author} {\bibfnamefont {J.}~\bibnamefont {Smith}}, \bibinfo
  {author} {\bibfnamefont {P.}~\bibnamefont {Richerme}}, \bibinfo {author}
  {\bibfnamefont {B.}~\bibnamefont {Neyenhuis}}, \bibinfo {author}
  {\bibfnamefont {P.~W.}\ \bibnamefont {Hess}}, \bibinfo {author}
  {\bibfnamefont {J.}~\bibnamefont {Zhang}}, \ and\ \bibinfo {author}
  {\bibfnamefont {C.}~\bibnamefont {Monroe}},\ }\href {\doibase
  10.1103/physreva.94.042308} {\bibfield  {journal} {\bibinfo  {journal}
  {Physical Review A}\ }\textbf {\bibinfo {volume} {94}} (\bibinfo {year}
  {2016}),\ 10.1103/physreva.94.042308}\BibitemShut {NoStop}%
\bibitem [{\citenamefont {Meekhof}\ \emph {et~al.}(1996)\citenamefont
  {Meekhof}, \citenamefont {Monroe}, \citenamefont {King}, \citenamefont
  {Itano},\ and\ \citenamefont {Wineland}}]{Meekhof1996}%
  \BibitemOpen
  \bibfield  {author} {\bibinfo {author} {\bibfnamefont {D.~M.}\ \bibnamefont
  {Meekhof}}, \bibinfo {author} {\bibfnamefont {C.}~\bibnamefont {Monroe}},
  \bibinfo {author} {\bibfnamefont {B.~E.}\ \bibnamefont {King}}, \bibinfo
  {author} {\bibfnamefont {W.~M.}\ \bibnamefont {Itano}}, \ and\ \bibinfo
  {author} {\bibfnamefont {D.~J.}\ \bibnamefont {Wineland}},\ }\href {\doibase
  10.1103/physrevlett.76.1796} {\bibfield  {journal} {\bibinfo  {journal}
  {Physical Review Letters}\ }\textbf {\bibinfo {volume} {76}},\ \bibinfo
  {pages} {1796–1799} (\bibinfo {year} {1996})}\BibitemShut {NoStop}%
\bibitem [{\citenamefont {Akaike}(1973)}]{Akaike1973}%
  \BibitemOpen
  \bibfield  {author} {\bibinfo {author} {\bibfnamefont {H.}~\bibnamefont
  {Akaike}},\ }\href@noop {} {\emph {\bibinfo {title} {Information theory and
  an extension of the maximum likelihood principle}}}\ (\bibinfo  {publisher}
  {Akad\'emiai Kiad\'o},\ \bibinfo {year} {1973})\ pp.\ \bibinfo {pages}
  {267--281},\ \bibinfo {note} {2nd International Symposium on Information
  Theory, edited by B. N. Petrov and F. Cs\'aki}\BibitemShut {NoStop}%
\end{thebibliography}%

\section*{APPENDICES}

\appendix

\section{Noise model derivations}

In the following, we derive the mean and variance of the fidelity $\mathcal{F}$ of Eq.~\ref{eq:fidelity_base} under various noise mechanisms. We first consider the noise-averaged fidelity, $\tilde{\mathcal{F}} = \langle \mathcal{F} \rangle_M$, where $\langle \cdot \rangle_M$ is the ensemble average taken over $M \rightarrow \infty$ noise realisations. We then consider the mean and variance of the noise-averaged fidelity, $\mathbb{E}[\tilde{\mathcal{F}}]_\Phi$ and $\mathbb{V}[\tilde{\mathcal{F}}]_\Phi$, taken with respect to all circuit realisations $\Phi$ of fixed length.

From Eq.~\ref{eq:fidelity_base}, the mean of the noise-averaged fidelity is
\begin{align} \label{eq:fidelity_moment}
   \mathbb{E}[\tilde{\mathcal{F}}] & = \mathbb{E}[ \langle  e^{-|\alpha_\epsilon|^2} \rangle_M ]_{\Phi}  \nonumber \\
   & = \sum_{k=0}^{\infty} \frac{(-1)^k}{k!}\mu_k,
\end{align}
where $\mu_k$ is the $k^\mathrm{th}$ moment of the distance $|\alpha_\epsilon|^2$, 
\begin{equation} \label{eq:moments}
    \mu_k = \mathbb{E}[\langle (|\alpha_\epsilon|^2)^k \rangle_M]_{\Phi}. 
\end{equation}
In what follows, we derive the moments of the distance $|\alpha_\epsilon|^2$ to evaluate the means of the fidelity. We find that the distributions $|\alpha_\epsilon|^2$ are well-approximated by Gamma distributions, $\Gamma(a, b)$, where $a$ is the shape parameter and $b$ is rate parameter. The moments of a Gamma distributed random variable $X\sim \Gamma(a, b)$ can conveniently be calculated with
\begin{equation} \label{eq:moments_general}
    \mu_k = \mathbb{E}[X^k] = b^k \frac{\Gamma(a + k)}{\Gamma(a)}.
\end{equation}

In particular, we consider a $\Gamma$-distribution with $a=1$, which will often appear in the following subsections. In this case, the moments of Eq.~\ref{eq:moments_general} become
\begin{equation}
    \mu_k = b^k k!,
\end{equation}
and the mean of Eq.~\ref{eq:fidelity_moment} simplifies to
\begin{equation} \label{eq:mean_fidelity_simplified}
    \mathbb{E}[\tilde{\mathcal{F}}] = \sum_{k=0}^{\infty} (-b)^k = \frac{1}{1 + b}.
\end{equation}

\subsection{Heating} 
\label{app:heating_noise_model}

The effects of heating are investigated by considering random displacements of $\epsilon_j$ that shift the bosonic mode during each step of the BRB protocol. Each random shift is sampled from $\epsilon_j \sim \mathcal{N}_\mathcal{C}(0, \sigma_\mathrm{h}^2)$, where $\mathcal{N}_\mathcal{C}$ is a complex normal distribution. The parasitic displacement is given by 

\begin{equation} \label{eq:alpha_eps_heating_app}
    \alpha_\epsilon = \sum_{j=0}^{J-1} \epsilon_j.
\end{equation}

Through the central limit theorem, $\alpha_\epsilon \sim \mathcal{N}_\mathcal{C}(0, J \sigma_\mathrm{h}^2)$. To calculate the parasitic distance $|\alpha_\epsilon|^2$, we note that the modulus square of a random variable $X \sim \mathcal{N}_\mathcal{C}(0, \sigma^2)$ is Gamma-distributed with $|X|^2 \sim \Gamma(1, \sigma^2)$. With this, we find that $|\alpha_\epsilon|^2 \sim \Gamma(a, b)$ is Gamma-distributed, with $a = 1$ and $b = J\sigma^2_\mathrm{h}$. We can finally calculate the mean of the noise-averaged fidelity from Eq.~\ref{eq:mean_fidelity_simplified},

\begin{align}
    \mathbb{E}[\tilde{\mathcal{F}}] & = \sum_{k=0}^\infty (-J \sigma_\mathrm{h}^2)^k = \frac{1}{1 + J \sigma_\mathrm{h}^2}.
\end{align}

We can express this last expression as a function of a heating rate, $\gamma_\mathrm{h}$, in units $\mathrm{quanta}/s$. In trapped ion systems, $\gamma_\mathrm{h}$ is often measured and used as a figure of merit. To this end, we follow the derivations of \cite{Savard1997, James1998b, rasmusson2024} and express $\alpha_\epsilon$ as
\begin{equation}
    \alpha_\epsilon = \frac{i e}{\sqrt{2 m \omega \hbar}} \int_{0}^{J \Delta\tau} E(t) e^{i\omega t} dt,
\end{equation}
where $e$ is the charge, $m$ the mass and $\omega$ is the motional frequency. $E(t)$ is a zero-mean noisy electric field and causes random displacements. The shifts $\alpha_\epsilon$ follow a complex normal distribution, with mean 

\begin{align}
    \mathbb{E}[\alpha_\epsilon] = \frac{i e}{\sqrt{2 m \omega \hbar}} \int_{0}^{J \Delta\tau} \mathbb{E}[E(t)] e^{i\omega t} dt = 0,
\end{align}
and variance 
\begin{align}
    \mathbb{V}[\alpha_\epsilon] & = \mathbb{E}[|\alpha_\epsilon|^2] \nonumber \\ 
    & = \frac{ e^2}{2 m \omega \hbar} \int_{0}^{J \Delta\tau} \int_{0}^{J \Delta\tau} \mathbb{E}[E(t) E(t')] e^{i\omega (t - t')} dt dt'.
\end{align}

Using the Wiener-Khinchin theorem, we express the variance as a function of the spectral density,  $S_E(\omega) = 2 \int^\infty_{-\infty} e^{i \omega \tau} \langle E(t) E(t + \tau)\rangle d\tau$,
\begin{align}
    \mathbb{V}[\alpha_\epsilon] & = \frac{e^2J\Delta\tau}{2m\omega\hbar} \int^{+\infty}_{-\infty} e^{i\omega\tau} \langle E(t)E(t + \tau)\rangle d\tau \nonumber \\
    & = \frac{e^2 J \Delta \tau S_E(\omega)}{4 m \hbar \omega}\nonumber \\
    & = \gamma_\mathrm{h} J \Delta\tau,
    \label{eq:var_alpha_eps_heating_rate}
\end{align}
where the heating rate is $\gamma_\mathrm{h} = \frac{e^2}{4 m \hbar \omega}S_E(\omega)$. Finally, relating the variance of Eq.~\ref{eq:var_alpha_eps_heating_rate} with the variance calculated from Eq.~\ref{eq:alpha_eps_heating_app}, $\mathbb{V}[\alpha_\epsilon] = \gamma_\mathrm{h}J\Delta\tau = J\sigma_\mathrm{h}^2$, we find the relation $\sigma^2_\mathrm{h} = \gamma_\mathrm{h}\Delta\tau$.

\subsection{Dephasing}
\label{app:dephasing_noise_model}

\begin{figure}[t]
    \centering
    \includegraphics[]{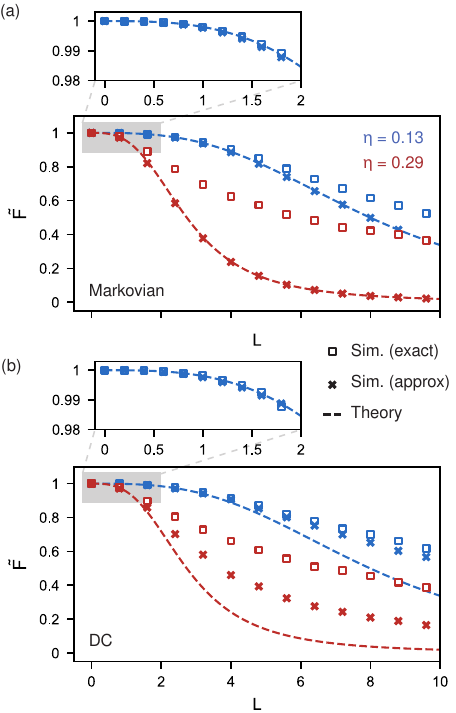}
    \caption{Validation of analytical error model for dephasing under Markovian noise (a) and quasi-static DC noise (b). The fidelity is averaged over $N=100$ circuit repetitions and $M=500$ noise realisations. Exact simulations (squares) are obtained from Eq.~\ref{eq:alpha_eps_dephasing_exact}. Approximate simulations (crosses) take the first order expansion of the exponential, $e^{- i \epsilon_\delta(t) t} \rightarrow 1 - i \epsilon_\delta(t) t$ (see Eq.~\ref{eq:alpha_eps_dephasing_app}). Analytical theory models (dashed line) are plotted from Eq.~\ref{eq:fidelity_model_dephasing_markov}. We consider noise strengths of $\sigma_\delta/2\pi = \SI{0.2}{kHz}$ (blue) and $\sigma_\delta/2\pi = \SI{1}{kHz}$ (red). The Rabi frequency is $\Omega/2\pi = \SI{1.65}{kHz}$ and the step sizes are $|\alpha_0| = 0.1$. Insets plot the fidelity in a smaller range of lengths, $L$.}
    \label{fig:theory_validation_dephasing}
\end{figure}

We now derive the expectation value and variance of the fidelity under dephasing. We consider a generalized noisy control Hamiltonian, 
\begin{equation}
    H_\mathrm{c} = \frac{\Omega}{2}(1 + \epsilon_\Omega(t))\hat{a}e^{-i (\phi(t) + \epsilon_\phi(t) + \epsilon_\delta(t) t)} + \mathrm{h.c.}
\end{equation}
which results in a noisy trajectory
\begin{align} \label{eq:noisy_alpha_tot_app}
    \tilde{\alpha}_\mathrm{tot} = & - \frac{i\Omega}{2} \sum_{j=0}^{J-1}  e^{- i \phi_j} \nonumber \\ 
    & \times \int_{j \Delta \tau}^{(j+1)\Delta \tau} (1 
+ \epsilon_\Omega(t)) e^{- i \epsilon_\delta(t) t} e^{- i \epsilon_\phi(t)}  dt,
\end{align}
where $\phi_j$ is randomly sampled from $\Phi$ to enact randomized displacements. $\epsilon_\Omega(t)$ are fractional amplitude fluctuations, $\epsilon_\delta(t)$ are frequency fluctuations of the harmonic oscillator and $\epsilon_\phi(t)$ are phase fluctuations. The ideal trajectory, $\alpha_\mathrm{tot}$, is retrieved by setting $\epsilon_\Omega(t) = \epsilon_\delta(t) = \epsilon_\phi(t) = 0$. Plugging Eq.~\ref{eq:noisy_alpha_tot_app} into Eq.~\ref{eq:alpha_epsilon}, we find that the parasitic displacement is
\begin{align} \label{eq:alpha_eps_app}
    \tilde{\alpha}_\epsilon = & - \frac{i\Omega}{2} \sum_{j=0}^{J-1}  e^{- i \phi_j} \nonumber \\ 
    & \times \int_{j \Delta \tau}^{(j+1)\Delta \tau} \left((1 
+ \epsilon_\Omega(t)) e^{- i \epsilon_\delta(t) t} e^{- i \epsilon_\phi(t)} - 1 \right)  dt,
\end{align}

In what follows, we separately consider the effects of frequency fluctuations (which we report in the main text), amplitude fluctuations and phase fluctuations.

\subsubsection{Frequency fluctuations}

We retrieve the parasitic displacement under frequency fluctuations by setting $\epsilon_\Omega(t) = \epsilon_\phi(t) = 0$, resulting in 
\begin{align} \label{eq:alpha_eps_dephasing_exact}
    \alpha_\epsilon & = - \frac{i \Omega}{2} \sum_{j=0}^{J-1} e^{-i \phi_j} \int_{j\Delta\tau}^{(j+1)\Delta\tau} (e^{- i \epsilon_\delta(t) t} - 1) dt.
\end{align}
The following derivations are simplified by expanding the exponential of the noisy fluctuations up to the first order, such that
\begin{align} \label{eq:alpha_eps_dephasing_app}
    \alpha_\epsilon & =  \frac{\Omega}{2} \sum_{j=0}^{N-1} e^{-i \phi_j} \int_{j\Delta\tau}^{(j+1)\Delta\tau} \epsilon_\delta(t) t dt + \mathcal{O}(\epsilon_\delta(t)^2t^2).
\end{align}
Higher order terms are negligible in the limit $\epsilon_\delta(t) t \ll 1$. 
We first derive the mean of the noise-averaged fidelity under uncorrelated Markovian and correlated DC noise. Markovian noise has a correlation length $\mathcal{M}_n = 1$ and is considered by replacing $\epsilon_\delta(t) \rightarrow \epsilon_{\delta, j}$ in Eq.~\ref{eq:alpha_eps_dephasing_app}. DC noise has a correlation length $\mathcal{M}_n = J$ and is modelled by replacing $\epsilon_\delta(t) = \epsilon_\delta$. In both cases, $\epsilon_\delta$ and $\epsilon_{\delta, j}$ are identical and independently distributed and sampled from a normal distribution $\mathcal{N}(0, \sigma_\delta^2)$ where $\sigma_\delta^2$ is the variance.

We find that, under both noise correlations, the parasitic displacement is $\Gamma$-distributed with $|\alpha_\epsilon|^2 \sim \Gamma(1, b)$, where $b = \frac{\Omega^2 \Delta\tau^4 \sigma_\delta^2 J^3}{12}$. Using Eq.~\ref{eq:mean_fidelity_simplified}, the mean of the noise-averaged fidelity becomes
\begin{align} \label{eq:fidelity_model_dephasing_markov}
    \mathbb{E}[\tilde{\mathcal{F}}] & = \frac{1}{1 + \Omega^2 \Delta\tau^4 \sigma_\delta^2 J^3/12}.
\end{align}
This final expression corresponds to Eq.~\ref{eq:fidelity_dephasing} of the main text after replacing $\eta = (4 |\alpha_0| \sigma_\delta^2 / 3\Omega^2)^{1/3}$.

We verify the validity of the analytical error model of Eq.~\ref{eq:fidelity_model_dephasing_markov} by comparing it to numerical simulations of the exact and approximate models of the parasitic displacements of Eq.~\ref{eq:alpha_eps_dephasing_exact} and Eq.~\ref{eq:alpha_eps_dephasing_app} (see Fig.~\ref{fig:theory_validation_dephasing}). We first observe that, for small errors $\eta \sim 0.1$ and short lengths $L < 2$, both the exact and approximate models are in great agreement with the analytical model (see insets). This suggests that the higher order terms $\mathcal{O}(\epsilon_\delta(t)^2t^2)$ of Eq.~\ref{eq:alpha_eps_dephasing_app} are negligible and the approximation is well motivated in this regime. However, for larger $L$, we observe that the exact results (squares) deviate from the analytical model (dashed line). Moreover, the discrepancy is more apparent for greater $\eta$ (red).

The variance of the fidelity under dephasing noise is calculated as
\begin{align}
    \mathbb{V}[\mathcal{\tilde{F}}] & = \mathbb{E}[\tilde{\mathcal{F}}^2] - \mathbb{E}[\tilde{\mathcal{F}}]^2.
\end{align}
For both DC and Markovian noise, we find 
\begin{align}
    \mathbb{E}[\tilde{\mathcal{F}}^2] \approx \frac{1}{1 + 2(\eta L)^3},
\end{align}
With this, the variance becomes
\begin{align} \label{eq:variance_th}
    \mathbb{V}[\mathcal{\tilde{F}}] = \frac{\mathbb{E}[\tilde{\mathcal{F}}](1 - \mathbb{E}[\tilde{\mathcal{F}}])^2}{2 - \mathbb{E}[\tilde{\mathcal{F}}]}
\end{align}

\begin{figure}[t]
    \centering
    \includegraphics[]{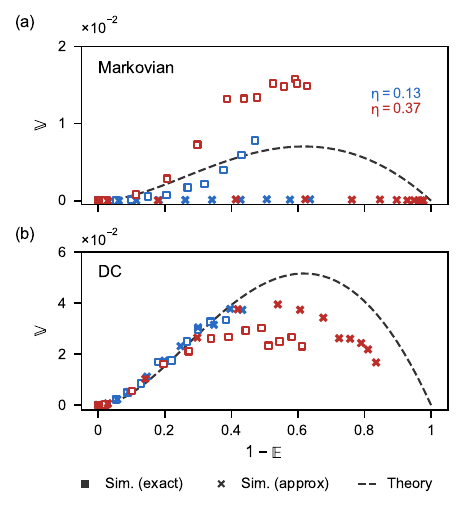}
    \caption{Validation of analytical variance model for dephasing under (a) Markovian noise and (b) quasi-static DC noise. Simulation parameters are identical to the results of Fig.~\ref{fig:theory_validation_dephasing}. Exact variance (squares, Eq.~\ref{eq:alpha_eps_dephasing_exact}) and approximate variance (crosses, Eq.~\ref{eq:alpha_eps_dephasing_app}) are compared to the analytical variance (dashed line) of Eq.~\ref{eq:variance_th}. The latter is scaled by $\mathcal{C}$, which is obtained from fits to numerical simulations.}
    \label{fig:theory_validation_variance_dephasing}
\end{figure}

This analytical expression of the variance is compared to exact and approximate models numerically simulated from Eq.~\ref{eq:alpha_eps_dephasing_exact} and Eq.~\ref{eq:alpha_eps_dephasing_app}, respectively (see Fig.~\ref{fig:theory_validation_variance_dephasing}). The exact and approximate variances agree well for small errors, $\eta \sim 0.1$ in the case of DC noise, but curiously continue to disagree for Markovian noise. This is likely due to non-trivial correlations that were ignored in calculating the variance and which do not appear for DC noise. We also find a discrepancy between the exact and approximate variances for larger errors, $\eta \sim 0.3$. Similarly to the results of Fig.~\ref{fig:theory_validation_dephasing}, this difference is due to higher order terms in Eq.~\ref{eq:alpha_eps_dephasing_app}, $\mathcal{O}(\epsilon_\delta(t)^2t^2)$, that were neglected. We further observe that the exact variances qualitatively follow the analytical model of Eq.~\ref{eq:variance_th}, within some scaling factor. Further discrepancies are most likely due to approximations made throughout the derivations, which ignore higher-order correlations. As these correlations are difficult to study for higher order terms, we content ourselves with scaling the analytical variance, $\mathbb{V} \rightarrow \mathcal{C} \mathbb{V}$, resulting in a semi-empirical model. The scaling factor $\mathcal{C}$ is extracted from fits to numerical simulations, which yields $\mathcal{C} = 0.071$ for Markovian noise and $\mathcal{C} = 0.572$ for DC noise.

\subsubsection{Amplitude fluctuations}

We here consider amplitude noise, and set $\epsilon_\delta(t) = \epsilon_\phi(t)=0$ in Eq.~\ref{eq:alpha_eps_app} resulting in
\begin{align} \label{eq:alpha_eps_amplitude_noise}
    \alpha_\epsilon = &- \frac{i \Omega}{2} \sum_{j=0}^{J-1} e^{-i\phi_j}\int^{(j+1)\Delta\tau}_{j\Delta\tau} \epsilon_\Omega(t) dt.
\end{align}
 Similarly to frequency fluctuations, we consider two types of correlations: uncorrelated Markovian noise and correlated DC noise. Under Markovian noise, we set $\epsilon_\Omega(t) \rightarrow \epsilon_{\Omega, j}$ in Eq.~\ref{eq:alpha_eps_amplitude_noise}, whereas, under DC noise, we set $\epsilon_\Omega(t) \rightarrow \epsilon_\Omega$. Here, $\epsilon_\Omega$ and $\epsilon_{\Omega, j}$ are i.i.d. and normally distributed from $\mathcal{N}(0, \sigma_\Omega^2)$ with variance $\sigma_\Omega^2$. 

Under both noise correlations, we find that the parasitic distance is $\Gamma$-distributed, $|\alpha_\epsilon|^2 \sim \Gamma(a, b)$, with $a = 1$ and $b = \Omega^2 \Delta\tau^2 \sigma^2_\Omega J / 4$. From Eq.~\ref{eq:mean_fidelity_simplified}, the mean of the noise-averaged fidelity is then
\begin{align}
    \mathbb{E}[\tilde{\mathcal{F}}] & = \frac{1}{1 + \Omega^2 \Delta\tau^2 \sigma_\Omega^2 J/4},
\end{align}
which can be expressed as 
\begin{align}
    \mathbb{E}[\tilde{\mathcal{F}}] = \frac{1}{1 + \eta L},
\end{align}
where $\eta = |\alpha_0| \sigma_\Omega^2$.

\subsubsection{Phase noise}

We here consider fluctuations of the phase by setting $\epsilon_\delta(t) = \epsilon_\Omega(t) = 0$ in Eq.~\ref{eq:alpha_eps_app}, for which the parasitic displacements are 
\begin{align}
    \alpha_\epsilon = - \frac{i \Omega}{2} \sum_{j=0}^{J-1}e^{-i \phi_j} \int_{j \Delta \tau}^{(j+1) \Delta\tau} (e^{- i \epsilon_\phi(t)} - 1) dt.
\end{align}
We consider small fluctuations of $\epsilon_\phi(t)$ around zero. In this limit, the parasitic displacements can be approximated as 
\begin{align}
    \alpha_\epsilon = \frac{\Omega}{2} \sum_{j=0}^{J-1}e^{-i \phi_j} \int_{j \Delta \tau}^{(j+1) \Delta\tau} \epsilon_\phi(t) dt.
\end{align}
We observe that this expression is nearly identical to the parasitic displacement under amplitude noise of Eq.~\ref{eq:alpha_eps_amplitude_noise}. Following the same derivations, we obtain the following decay,
\begin{align}
    \mathbb{E}[\tilde{\mathcal{F}}] = \frac{1}{1 + \eta L},
\end{align}
where $\eta = |\alpha_0| \sigma_\phi^2$.

\section{Displacements with state-dependent forces}
\label{app:displacements}

State-independent displacements of the form $\hat{\mathcal{D}}(\alpha)$ can be implemented by mapping the qubit in and out of a basis that is orthogonal to the SDF interaction. This is done by aligning the qubit state along the Bloch sphere's $Y$ axis with $\pi/2$ pulses at the beginning and end of the BRB protocol. In practice, however, we omit the $\pi/2$ pulses to avoid complications from unequal Stark shifts between the displacements and the pulses~\cite{Lee2016}. From numerical simulations, we find that the behaviour of the BRB protocol is not impacted by the state dependency of the light-atom interaction, provided that the fields are resonant and that the qubit coherence time is sufficiently long. To this end, we perform frequent interleaved calibrations during BRB sequences; furthermore, we measure a qubit coherence time of $T_2 \sim \SI{1}{s}$, sufficiently longer than the duration taken for each BRB sequence.

\section{Reconstruction measurements}
\label{app:reconstruction_measurements}

The noise-averaged fidelity, $\tilde{\mathcal{F}}$, is experimentally retrieved by applying a $\pi$-pulse on the red-sideband transition. A measurement of the qubit then yields the ideal probability
\begin{equation} \label{eq:rsb_pulse_prob}
    P_{\ket{\downarrow}} = \frac{1}{2} + \frac{1}{2} \sum_{n=0}^{\infty} P_n \cos(\pi \sqrt{n}),
\end{equation}
where $P_n$ is the probability distribution in the Fock basis. Throughout section~\ref{sec:error_model} of the main text, it was shown that the various noise mechanisms under consideration cause the bosonic mode to be displaced by $\alpha_\epsilon$ from the origin. The probability distribution $P_n$ after a single noise realisation can therefore be described by that of a displaced state,
\begin{equation} \label{eq:prob_coherent}
    P_n = e^{- |\alpha_\epsilon|^2} \frac{(|\alpha_\epsilon|^2)^n}{n!}.
\end{equation}
In the limit of small errors, $|\alpha_\epsilon|^2 \ll 1$, and plugging Eq.~\ref{eq:prob_coherent} into Eq.~\ref{eq:rsb_pulse_prob}, the probability $P_{\ket{\downarrow}}$ becomes

\begin{equation} \label{eq:rsb_pulse_prob_approx}
    P_{\ket{\downarrow}} \approx 1 - |\alpha_\epsilon|^2.
\end{equation}

Finally, averaging Eq.~\ref{eq:rsb_pulse_prob_approx} over many noise realisations $M$ such that $\tilde{P}_{\ket{\downarrow}} = \langle P_{\ket{\downarrow}} \rangle_M$, we find that $ \tilde{P}_{\ket{\downarrow}} \approx \tilde{\mathcal{F}}$. We note that a better estimate of $\tilde{\mathcal{F}}$ could be obtained at the expense of more measurements, by applying, for example, the red-sideband for varying durations~\cite{Meekhof1996}.

\section{Model comparison for intrinsic hardware noise characterization}
\label{app:aic}

We use the Akaike information criteria (AIC) \cite{Akaike1973} in section~\ref{sec:hardware_characterization} to determine the most likely model corresponding to the experimental benchmarking data of Fig.~\ref{fig:exp_benchmarking}. Assuming that the residuals are identical and independently distributed, the AIC is determined from the residual sum of squares (RSS),

\begin{equation}
    \mathrm{AIC} = n \mathrm{log}(\frac{\mathrm{RSS}}{n}) + 2k,
\end{equation}

where $n$ is the number of data points and $k$ is the number of model parameters. In both heating and dephasing models, the only model parameter is the error rate, $\eta$, and we set $k=1$.

\end{document}